\documentclass[manuscript]{aastex}
\usepackage{graphicx}
\usepackage{longtable}
\usepackage{epsfig}
\usepackage{lscape}

\newcommand{\psr}{PSR~J2021+4026}
\newcommand{\G}{G78.2+2.1}
\begin{document}
\slugcomment{}

\shorttitle{Deep XMM observation of \psr\ \& $\gamma-$Cygni SNR}

\title{A detailed X-ray investigation of \psr\ and $\gamma-$Cygni supernova remnant}
\shortauthors{Hui et al.}

\author{C. Y. Hui\altaffilmark{1}, K. A. Seo\altaffilmark{1}, L. C. C. Lin\altaffilmark{2}, R. H. H. Huang\altaffilmark{2}, C. P. Hu\altaffilmark{3}, J. H. K. Wu\altaffilmark{2}, 
 L. Trepl\altaffilmark{4}, J. Takata\altaffilmark{5}, Y. Wang\altaffilmark{5}, Y. Chou\altaffilmark{3}, 
K. S. Cheng\altaffilmark{5} and A. K. H. Kong\altaffilmark{2}}

\altaffiltext{1}
{Department of Astronomy and Space Science, Chungnam National University, Daejeon, Republic of Korea}
\altaffiltext{2}
{Institute of Astronomy and Department of Physics,
National Tsing Hua University, Hsinchu, Taiwan }
\altaffiltext{3}{Graduate Institute of Astronomy, National Central University,
 Jhongli 32001, Taiwan}
\altaffiltext{4}
{Astrophysikalisches Institut und Universit\"{a}ts-Sternwarte,
Universit\"{a}t Jena, Schillerg\"{a}$\beta$chen 2-3, 07745 Jena,
Germany}
\altaffiltext{5}
{Department of Physics, University of Hong Kong, Pokfulam Road, 
Hong Kong}

\email{cyhui@cnu.ac.kr}


\begin{abstract}
We have investigated the field around the radio-quiet $\gamma$-ray pulsar, \psr, 
with a $\sim140$~ks \emph{XMM-Newton} observation and a $\sim56$~ks archival \emph{Chandra} data. 
Through analyzing the pulsed spectrum, we show that the X-ray pulsation is purely thermal in nature
which suggests the pulsation
is originated from a hot polar cap with $T\sim3\times10^{6}$~K on the surface of a rotating neutron star. 
On the other hand, the power-law component that dominates the pulsar emission in the hard band is originated from 
off-pulse phases, which possibly comes from a pulsar wind nebula. In re-analyzing the 
\emph{Chandra} data, we have confirmed the presence of bow-shock nebula which extends from the pulsar 
to west by $\sim10$~arcsec. The orientation of this nebular feature suggests that the pulsar is probably 
moving eastward which is consistent with the speculated proper motion by extrapolating from  
the nominal geometrical center of the supernova remnant (SNR) \G\ to the current pulsar position. 
For \G, our deep \emph{XMM-Newton} observation also enables a study of the central region and part of the 
southeastern region with superior photon statistics. The column absorption derived for the SNR is comparable with 
that for \psr, which supports their association. The remnant emission in both examined regions are in an 
non-equilibrium ionization state. Also, the elapsed time of both regions after shock-heating is apparently shorter 
than the Sedov age of \G. 
This might suggest the reverse shock has reached the center not long ago. 
Apart from \psr\ and \G, we have also serendipitously detected an X-ray flash-like event
XMM~J202154.7+402855 from this 
\emph{XMM-Newton} observation. 
\end{abstract}

\section{INTRODUCTION}
The Cygnus region, which locates at $\sim80^{\circ}$ from the Galactic center, is one of the 
most complex $\gamma-$ray structure on the Galactic plane. Many interesting sources are found 
to reside in this region, including OB association, microquasar, supernova remnants and rotation-powered 
pulsars. Since the launch of the {\it Fermi} Gamma-Ray Space Telescope, many new $\gamma-$ray pulsars 
have been uncovered in the Cygnus region either in blind searches or by folding the $\gamma-$ray data with 
the timing ephemeris determined through radio observations (Abdo et al. 2013). Among them, the most intriguing 
one is \psr, which belongs to the growing class of radio-quiet $\gamma$-ray pulsars. 

\psr\ is a bright $\gamma$-ray source which was detected at a significance $>10\sigma$ 
with only the first 3 months LAT data (Abdo et al. 2009a 0FGL). Its timing ephemeris was subsequently 
reported (Abdo et al. 2009b, 2010). It has a rotation period and first 
period derivative of $P=265$~ms and $\dot{P}=5.48\times10^{-14}$~s~s$^{-1}$ respectively, which 
imply a spin-down age of $\sim77$~kyr, surface dipolar magnetic field strength of $\sim4\times10^{12}$~G 
and a spin-down power of $\sim10^{35}$~erg~s$^{-1}$. 

Recently, the long-term $\gamma-$ray monitoring of \psr\ has discovered its $\gamma-$ray flux at energies 
$>100$~MeV has suddenly decreased by $\sim18\%$ near MJD~55850 (Allafort et al. 2013). This flux jump 
was accompanied by the change in the $\gamma-$ray pulse profile and the spin-down rate. These make 
\psr\ to be the first variable $\gamma-$ray pulsar has ever been observed. 

Despite the efforts devoted in searching for its radio 
counterpart, no radio pulsar associated with \psr\ has been detected so far (Becker et al. 2004; 
Trepl et al. 2010; Ray et al. 2011). 
Without radio detection, observations in X-ray regime is particularly important 
for constraining the emission nature of radio-quiet $\gamma$-ray pulsars. 
Soon after the detection of \psr\ has been reported by Abdo et al. (2009a), we 
have searched for its X-ray counterpart by using all the avaliable archival data (Trepl et al. 2010). Within 
the 95\% confidence circle of \psr\ at that time (0FGL~J2021.5+4026; cf. Abdo et al. 2009a), we reported 
the identification of the X-ray source 2XMM~J202131.0+402645 / CXOU 202130.55+402646.9 
from the {\it XMM-Newton} (ObsID: 150960801) and {\it Chandra} (ObsID: 5533) data 
as the possible counterpart of \psr. This source was found to be the only non-variable X-ray object without 
any optical/IR counterpart within the $\gamma$-ray error circle. The association between the pulsar 
and this source is reinforced by the fact that the X-ray position is consistent with the optimal $\gamma$-ray 
timing solution (Trepl et al. 2010). 
These results have been confirmed by a follow-up 56~ks {\it Chandra} observation for a dedicated investigation 
(ObsID 11235; Weisskopf et al. 2011). 
This observation 
also enables the authors to examine the X-ray spectrum of this promising counterpart of \psr\ and found that it 
possibly contains both thermal and non-thermal contributions. 

Although all the aforementioned investigations strongly suggest the association between 2XMM~J202131.0+402645 
and \psr, 
the physical connection between this X-ray source and the $\gamma$-ray pulsar 
could not be confirmed unambiguously until the X-ray pulsation was recently discovered by us
with a deep {\it XMM-Newton} observation (Lin et al. 2013). The consistency between the detected X-ray periodicity and 
the $\gamma-$ray pulsation at the same epoch has eventually nailed down the long-sought connection between 2XMM~J202131.0+402645
and \psr. In this paper, we reported a further X-ray analysis of \psr\ by disentangling its pulsed and unpulsed components. 
Also, by re-examining archival on-axis \emph{Chandra} observation, we have searched for the possible pulsar wind nebula around \psr. 

Apart from \psr, our {\it XMM-Newton} observation also covers the central region and part of the southeastern 
rim of the supernova remnant 
(SNR) $\gamma-$Cygni (\G) which has been suggested to be associated with \psr\ (Trepl et al. 2010). 
\G\ is rather extended. Its radio and X-ray shells have a size of $\sim1^{\circ}$ (Leahy, Green \& Ranasinghe 2013). 
The X-ray emission from \G\ has been investigated several times with \emph{ROSAT}, \emph{ASCA} and \emph{Chandra} 
(Lozinskaya et al. 2000; Uchiyama et al. 2002; Aliu et al. 2013; Leahy et al. 2013). 
The most recent X-ray imaging spectroscopic analysis of \G\ is reported by Leahy et al. (2013). 
By using the archival \emph{Chandra} data, the authors have examined the diffuse X-ray emission from the northern part and the central 
region of \G. While the column absorption for these two spatial regions are found to comparable, the plasma temperature of 
the central region, $\sim10^{7}$~K, is suggested to be higher. However, the uncertainties of the spectral parameters are too large 
for drawing a firm conclusion. 
This is can be ascribed to the low surface brightness of the central region (see Fig.~4 in Leahy et al. 2013)
and the relatively inferior collecting power of \emph{Chandra}. 

For the southeastern rim, it contributes $\sim60\%$ of the radio flux of this SNR which is known as DR4 (Downes \& Rinehart 
1966). While this is the brightest part in radio, it is very dim in X-ray regime (see Fig. 1 in Uchiyama et al. 2002).
This part has been excluded in the updated investigation by Leahy et al. (2013), as there is no existing \emph{Chandra} data that
covers this region. With our deep {\it XMM-Newton} observation, we are able to provide tighter constraints on the emission 
nature of the central and southeastern part of \G\ and its possible association with \psr. 

\section{OBSERVATION \& DATA ANALYSIS}
\subsection{Observations \& Data Reduction}
Our deep {\it XMM-Newton} observation of the field around \psr\ started on 2012 April 11 with a total on-time 
of 135.8~ks (ObsID: 0670590101; PI: Hui). {\bf E}uropean {\bf P}hoton
{\bf I}maging {\bf C}amera (EPIC) was used throughout this investigation. 
The PN CCD was operated in small-window mode with a medium filter to block optical stray light.
All the events recorded by PN camera are time-tagged with a temporal resolution of 5.7 ms, 
which enable us to examine the spectral properties at different rotational phases for the first time. 
On the other hand, the MOS1/2 CCDs were operated in
full-window mode with a medium filter in each camera, which provide us with a large field-of-view  
(15' in radius) for a deep search of X-ray point sources as well as the diffuse X-ray emission 
from SNR \G. 
The median satellite boresight pointing during this observation is 
RA=$20^{\rm h}21^{\rm m}30.56^{\rm s}$ Dec=$+40^{\circ}26^{'}46.8^{''}$ (J2000), which is the position 
of 2XMM~J202131.0+402645 determined by Trepl et al. (2010). 

With the most updated instrumental calibration, we
generated the event lists from the raw data obtained from all EPIC instruments
with the tasks \emph{emproc} and \emph{epproc} of the {\bf XMM}-Newton {\bf S}cience {\bf A}nalysis 
{\bf S}oftware (XMMSAS version 12.0.1). The event files were subsequently filtered for the energy range 
from 0.5 keV to 10 keV for all EPIC instruments and selected only those events for which the pattern was
between $0-12$ for MOS cameras and $0-4$ for the PN camera.
We further cleaned the data by accepting only the good times when the sky background was low and
removed all events potentially contaminated by bad pixels. After the filtering, the effective
exposures are found to be 85~ks, 72~ks and 77~ks for MOS1, MOS2, PN respectively.  

We have also re-analysed the archival {\it Chandra} data with \psr\ on-axis (Obs. ID: 11235, PI: Weisskopf) 
in order to constrain the evidence for the pulsar wind nebula (PWN). This observation has used the 
Advanced CCD Imaging Spectrometer (ACIS) with the aim-point on the back-illuminated CCD ACIS-S3 for 
an exposure of 56~ks. The major results of this observation have already been reported by 
Weisskopf et al. (2011). In our investigation, we focus on searching for the possible extended 
emission associated wtih \psr\ with the sub-arcsecond angular resolution of this data, which has not yet 
been fully explored by Weisskopf et al. (2011). By using the script \emph{chandra\_repro} provided 
in the {\bf C}handra {\bf I}nteractive {\bf A}nalysis {\bf O}bservation software (CIAO~4.3), we have reprocessed the data 
with CALDB (ver. 4.4.5). Since we aim for a high spatial resolution analysis, 
sub-pixel event repositioning has been applied during the data reprocessing in order 
to improve the positional accuracy of each event (cf. Li et al. 2004). We restricted the analysis of 
this ACIS data in an energy range of $0.3-8$~keV. 

\subsection{Spatial Analysis}
The X-ray color images as obtained by MOS1/2 and PN are shown in Figure~\ref{xmm_color} and Figure~\ref{pn_img} 
respectively (Red:0.5-1~keV; Green:1-2~keV; Blue:2-10~keV). In order to correct for the non-uniformity across the 
detector and the mirror vignetting, each image has been normalized by the exposure map 
generated by using the XMMSAS task \emph{eexpmap} for the corresponding detector. 
The exposure-corrected images have been adaptively smoothed so as to attain a minimum signal-to-noise 
ratio of 3. This deep observation allows us to search for new X-ray point sources in this field. 
In Figure~\ref{xmm_color}, point sources with various X-ray hardness 
can be seen. For determining their positions and count rates, we performed a source detection by 
using maximum likelihood fitting on MOS1, MOS2 and PN data individually with the aid of the XMMSAS task
\emph{edetect\_chain}. We set the detection threshold to be $4\sigma$. 
 
By visual inspection, we removed the weak sources which 
are potentially false detections from the source lists resulted from individual cameras. 
These include several sources close to the edge of the field-of-view (1 from MOS1, 1 from MOS2, 2 from PN)
as well as a few slightly extended sources coincide with the diffuse emission which are likely to be 
clumps of the supernova remnant (3 from MOS1 and 2 from MOS2). 
The screened lists are subsequently
correlated and merged by using the XMMSAS task \emph{srcmatch}. In case the position of a source obtained from 
two detections are consistent within their $3\sigma$ uncertainties, they are merged as a single entry. 
Their source properties are summarized in Table~1\footnote{Table~1 is slightly different from the standard 
source list given by {\it XAssist} in three aspects: (1) Multiple detections with positional coincidence are 
combined as single entry. (2) Potentially false detections are removed by visual inspection. (3) All the tabulated 
sources have signal-to-noise ratio $>$4.}
Including \psr\ (i.e. source 18), $42$ distinct point sources are detected within the 30' field-of-view (FoV) around 
the aim-point in this observation. Among all these X-ray point-like objects in this field, source 8 
(XMM~J202154.7+402855) is the brightest. A further analysis found that its X-ray flux is significantly variable which
worth a deeper investigation. The spectral and temporal analysis of XMM~J202154.7+402855 are reported in appendix. 

Besides the point source population, faint diffuse X-ray structures have also been seen in this observation. 
At the southeastern edge of the MOS1/2 image (Fig.~\ref{xmm_color}), extended soft emission is highlighted by 
the elliptical region. This region partially covers the structure R1 examined by Uchiyama et al. (2002). 
In the central region, an extended structure around \psr\ can be seen in both Fig.~\ref{xmm_color} and Fig~\ref{pn_img}. 
A close-up look of this feature as seen by the PN camera is shown in Figure~\ref{pn_img}. 
The solid-line ellipses in Fig.~\ref{xmm_color} and Figure~\ref{pn_img}
illustrate the source regions for extracting the spectra from the southeastern and the central parts of \G\ respectively (see \S2.4). 

Apart from reporting the discovery of X-ray pulsation, Lin et al. (2013) have also found that the phase-averaged 
spectrum can be described by a blackbody plus power-law model. One possible origin for the power-law component
is the PWN. Although this {\it XMM-Newton} observation provides a superior photon statistics for
spectral and timing analysis, its relatively wide point spread 
function (PSF) does not allow us to search for the possible compact PWN around the pulsar. 

In investigating the {\it Chandra} ACIS image, Weisskopf et al. (2011) have briefly mentioned a possible feature 
associated with \psr\ which may be indicative of a PWN. By fitting the X-ray image of \psr\ with a 
circular Gaussian plus a constant background, the authors concluded that it is consistent with being point-like and 
placed an upper-limit of its extent to be $\lesssim6"$. Although this may be useful in quantifying the extent 
of bright and symmetric nebula (e.g. plerionic emission associated with young pulsars like Crab), such method 
can overlook faint asymmetric extended feature. 
This motivates us to reexamine this archival data to characterize the properties of this feature in details. 
The adaptively-smoothed ACIS image of a $0.5'\times0.5'$ field around \psr\ is shown in 
Figure~\ref{cxc_pwn_img}. It clearly shows a nebula-like structure which extends to west from 
\psr. For the PWN associated with a fast-moving pulsar, the extended 
X-ray emission is typically aligned with the direction of proper motion (cf. Gaensler 2005). 
Although the proper motion of \psr\ is unknown, the orientation 
of its associated extended X-ray PWN indicates that it might be moving eastward. Assuming the birth place of  
\psr\ is not far away from the nominal center of \G\ given by Green (2009) (i.e. the white cross in 
Fig.~\ref{xmm_color}), we speculate the direction of the proper motion by extrapolating  
from the nominal remnant center to the current pulsar position. The speculated direction is illustrated by the 
white arrow in Figure~\ref{cxc_pwn_img}. It is interesting to notice that it deviates from the symmetric axis of the nebula by only 
$\sim5^{\circ}$. 

For quantifying the extent of such elongated compact nebula, we computed 
its brightness profile along its orientation (Hui et al. 2007, 2008, 2012). 
We estimated the counts in 18 consecutive boxes with a size of $1''\times10''$ from the raw image
with a pixel size of $0.5^{"}\times0.5^{"}$ along the extended feature (see the inset of Fig.~\ref{brightness}).
The observed brightness profile is shown in Figure~\ref{brightness}. 
To estimate the background level, we have sampled the source-free
regions around \psr\ within a $2'\times2'$ field-of-view. The average background level and its
$1\sigma$ uncertainties are indicated by the horizontal solid line and dotted lines 
respectively. The nebular feature apparently extends for $\sim10''$ to the west before it falls
to the estimated background level, which clearly exceeds the upper-limit placed by Weisskopf et al. (2011) through a 
symmetric Gaussian fitting. 

To further examine its emission nature, we have extracted the photons from this feature within a
box of $9'\times6'$ centered at
RA=$20^{\rm h}21^{\rm m}30.19^{\rm s}$ Dec=$+40^{\circ}26^{'}45.2^{''}$ (J2000). Only 25 counts are
collected from this observation which forbids any meaningful spectral analysis. However, we can still
estimate its X-ray color and compare with \psr. Following Trepl et al. (2010) and Weisskopf et al. (2011),
we divide the energy range into three bands:
soft ($S$: 0.5-1~keV), medium ($M$: 1-2~keV) and hard ($H$: 2-8~keV).
The X-ray colors of this extended feature are estimated to be $(H-S)/T=0.41\pm0.21$ and $M/T=0.59\pm0.16$.
In comparison with the X-ray colors of \psr, $(H-S)/T=0.02\pm0.03$ and $M/T=0.75\pm0.02$ 
(Weisskopf et al. 2011), the X-ray emission from the extended feature is apparently harder which 
possibly indicates its non-thermal nature. 

\subsection{Analysis of the pulsed X-ray emission from \psr}
\subsubsection{Pulsed spectrum of \psr}
From the phase-averaged spectral analysis, it has been shown that a single component model is not able to 
describe the observed data beyond $\sim3$~keV (Weisskopf et al. 2011; Lin et al. 2013). 
Statistically, an additional hard component is required at a confidence level 
$>99.995\%$. A blackbody (BB) plus power-law (PL) model fits the data reasonably well (cf. Fig.~3 in Lin et al. 2013). 
Although the PWN as seen by \emph{Chandra} is likely to contribute at least a part of this hard component, one cannot rule out the possiblity 
that the observed non-thermal X-rays are originated from the pulsar magnetosphere and thus have contributions to the observed 
pulsation. To distinguish these two scenarios, 
one has to determine in which rotational phases does this PL component dominate. This motivates us to examine 
the pulsed spectrum of \psr\ with the PN data. 

We extracted the source spectrum in a circular region of a 20$''$ radius centered at the X-ray position of \psr\ as 
determined by the source detection algorithm. The adopted extraction region corresponds to an encircled 
energy function of $\sim76\%$. 
For determining the pulse phase of each photon, their arrival times were firstly corrected to the solar system 
barycenter with the aid of XMMSAS task \emph{barycen} by using the JPL DE405 earth ephemeris. For assigning the 
pulse phase to each event, we adopted the temporal parameters as determined by Lin et al. (2013). 

We divided the rotational phases into two regimes, which are illustrated by the shaded regions in 
Figure~\ref{pulse_phase}. The phase intervals $0.2-0.7$ and $0.85-1.2$ are defined as the ``on-pulse" and 
``off-pulse" components respectively. Assuming the off-pulse component contributes a steady DC level across 
the entire rotational phase, 
the pulsed spectrum was then obtained by subtracting the off-pulse component from the on-pulse component. 
The response files were produced by the XMMSAS task \emph{rmfgen} and \emph{arfgen}. The spectrum is binned 
so as to have $>50$ counts per spectral bin. We used XSPEC 12.7.0 for all the spectral analysis reported in this work. 
The spectral fits were performed in 0.5-10~keV. 

The pulsed spectrum is found to be softer than the phase-averaged spectrum. 
Majority of the pulsed X-rays have energies $<3$~keV (see Figure~\ref{psr_spec}). 
At energies $>3$~keV, $\gtrsim72\%$ of the collected photons are contributed by the off-pulse component. 
We further found that the pulsed spectrum can be well described 
by a simple absorbed blackbody model ($\chi^{2}=12.86$ for 15 D.O.F.) without requiring additional components. 
The best-fit model yields a column density of 
$N_{\rm H}=(9.1^{+5.2}_{-3.5})\times10^{21}$~cm$^{-2}$, a temperature of $kT=0.23^{+0.06}_{-0.05}$~keV, a
blackbody emitting region with a radius of $R=500^{+1053}_{-282}d_{2}$~m, where $d_{2}$ represents the distance to \psr\ in unit 
of 2~kpc. As the uncertainty of $N_{\rm H}$ is large, we fixed it at the value derived from the phase-averaged analysis 
(Lin et al. 2013), $N_{\rm H}=7\times10^{21}$~cm$^{-2}$, and constrained the blackbody temperature and radius to be 
$kT=0.26^{+0.03}_{-0.02}$~keV and $R=318^{+101}_{-77}d_{2}$~m respectively ($\chi^{2}=13.51$ for 16 D.O.F.). 
For a conservative estimate, all the quoted errors of the spectral parameters
are $1\sigma$ for two parameters of interest (i.e. $\Delta\chi^{2}=2.30$ above the minimum). 

We have also attempted to perform a phase-resolved spectroscopy for each phase bin in Figure~\ref{pulse_phase} 
in order to investigate 
how does the spectral properties varies across the rotational phase. However, the photon statistic for 
individual phase bins is generally
too small for a constraining analysis. The high instrumental background has further exacerbated the situation. 
Therefore, such analysis
will not be further considered for this observation. 

\subsubsection{Multi-epoch X-ray spectral analysis of \psr.}
As mentioned in the introduction, \psr\ is the first variable radio-quiet $\gamma-$ray pulsar where its $\gamma-$ray 
flux as spin-down properties suddenly change around MJD~55850 (2011 October 16). 
Allafort et al. (2013) argued that the variability of the $\gamma-$ray 
pulsed emission is due to certain global change in the magnetosphere. Since the X-ray emission from the hot polar cap 
is resulted from the 
bombardment of the backflow current from the outergap (Cheng \& Zhang 1999), it is not unreasonable to speculate that 
the change in the $\gamma-$ray properties might be accompanied with the change in X-ray. 

In the context of the outergap model 
(Cheng \& Zhang 1999), the $\gamma$-ray luminosity is given by $L_{\gamma}\sim f^{3}\dot{E}$ where $f$ is the fractional size of 
the outergap. Therefore, the observed change in the $\gamma$-ray luminosity, $\delta L_{\gamma}/L_{\gamma}\sim18\%$ (Allafort et al. 2013),
implies the gap size changed by $\delta f/f\sim\delta L_{\gamma}/3L_{\gamma}\sim6\%$. Since our analysis suggests the pulsed X-ray emission 
is thermal, this can be produced through the polar cap heating by the return particle flux $\dot{N_{p}}=f\dot{N_{\rm GJ}}$ where 
$\dot{N_{\rm GJ}}$ is the Goldreich-Julian particle flux (Goldreich \& Julian 1969). The X-ray luminosity can thus be estimated by 
$L_{X}\sim\dot{N_{p}}E_{p}$ where $E_{p}$ is the typical particle energy deposited on the stellar surface. Therefore, the 
expected change in $L_{X}$ is found to be $\delta L_{X}/L_{X}\sim\delta\dot{N_{p}}/\dot{N_{p}}\sim\delta f/f\sim6\%$.  

Since the \emph{Chandra} (MJD~55435) and the
\emph{XMM-Newton} (MJD~56028) observations used in this study were performed before and after $\gamma-$ray flux jump, it is instructive to
examine if \psr\ exhibited any X-ray variability.
In order to investigate whether the X-ray spectral properties of \psr\ vary, 
we examined the phase-averaged spectra of \psr\ as obtained by \emph{Chandra} ACIS-S3 and all EPIC camera on \emph{XMM-Newton}. 
For \emph{Chandra}, we extracted the source spectrum from a circular region with a radius of 2" centered at the pulsar 
position. The background spectrum was sampled from an annular region with inner/outer radii of 2.5"/4" around the pulsar. 
For \emph{XMM-Newton}, we followed the same procedure adopted by Lin et al. (2013) in preparing the phase-averaged spectrum. 

We jointly fitted an absorbed BB+PL model to the \emph{Chandra} and 
\emph{XMM-Newton} spectra with the column absorption tied together throughout the analysis. 
In order to minimize the number of free parameters, we examined the variability of each spectral component one at a time.
First, with the PL component in different epoch tied together, 
we allowed the BB component of the spectrum in different
epoch to vary independently during 
the fitting. This yielded a column absorption of $N_{H}=6.4^{+0.8}_{-1.8}\times10^{21}$~cm$^{-2}$, a photon index of $\Gamma=1.5\pm0.8$ and 
a PL normalization of $3.6^{+6.8}_{-2.6}\times10^{-6}$~photons~keV$^{-1}$~cm$^{-2}$~s$^{-1}$ at 1 keV.
For the BB component, the best-fit temperature and emission radius for the epoch MJD~55435 are $kT=0.22\pm0.04$~keV and 
$R=350^{+359}_{-94}d_{2}$~m respectively. After the $\gamma-$ray flux change, the BB parameters are found to be 
$kT=0.24^{+0.04}_{-0.02}$~keV and $R=288^{+193}_{-27}d_{2}$~m in the epoch MJD~56028. 

For inspecting the possible variability of the PL component, the BB components of both spectra were tied. It 
yielded a column absorption of $N_{H}=6.4^{+0.8}_{-1.8}\times10^{21}$~cm$^{-2}$, a blackbody temperature of $kT=0.24\pm0.04$~keV 
and a emission radius of $R=298^{+22}_{-96}d_{2}$~m. The photon index and PL normalization before the $\gamma-$ray flux change are found to be 
$\Gamma=1.0^{+2.0}_{-1.0}$ and $1.4^{+11.1}_{-1.4}\times10^{-6}$~photons~keV$^{-1}$~cm$^{-2}$~s$^{-1}$ at 1 keV. The corresponding 
parameters after the $\gamma-$ray flux change are 
$\Gamma=1.8\pm0.8$ and $5.4^{+10.1}_{-3.7}\times10^{-6}$~photons~keV$^{-1}$~cm$^{-2}$~s$^{-1}$ at 1 keV.

Within the tolerence of the quoted statistical uncertainties, neither the pulsed thermal X-ray component nor the unpulsed non-thermal 
X-ray component are found to be variable in these two epochs. 

\subsubsection{Analysis of the X-ray pulse profile of \psr.}
The X-ray pulse profile can also be used to investigate the global properties of the neutron star. As the polar cap sweeps across 
our line-of-sight, modulation in soft X-ray regime can be seen (cf. Hui \& Cheng 2004; Pechenick et al. 1983). Since the gravity of a
neutron star is tremendous, the shape of thermal X-ray pulse profile is determined by the near-field spacetime curvature. 
The effect of gravity on the trajectory of emitted photons, which depends on the mass-to-radius ratio of the neutron star, 
must be considered in modeling the light curve. 

Following Pechenick et al. (1983), we simulated the X-ray pulse profile resulted 
from the general relativistic calculation and compare with the observational result. 
We choose our coordinates so that the observer is on the positive $z$-axis at $r=r_{0}$ where $r_{0}\rightarrow\infty$ 
(cf. Fig.~6 in Hui \& Cheng 2004). We described the stellar surface by angular spherical coordinates $\theta$ and $\phi$ 
where $\theta$ is measured from the $z$-axis defined above.  
For the photon emitted at an angle $\delta$ from the stellar surface, it will seem to the observer that they 
are emitted at an angle $\theta^{'}$ from the $z$-axis as a result of gravitational light bending. The relationship between 
$\theta$ and $\theta^{'}$ is given by:
\begin{equation}
\theta=\int_{0}^{\frac{GM}{Rc^{2}}}\left[\left(\frac{GM}{bc^{2}}\right)^{2}-\left(1-2u\right)u^{2}\right]^{-1/2}du,
\end{equation}
\noindent where $b=r_{0}\theta^{'}$ is the impact parameter of the photon and $u=GM/c^{2}r$. 

For a neutron star, $\frac{GM}{c^{2}R}$ must be less than 1/3. Therefore, a photon emitted from the surface that reaches the 
observer must have an impact parameter $b\leq b_{max}$ where $b_{max}=R\left(1-2GM/c^{2}R\right)^{-1/2}$ (Pechenick et al. 1983). 
The condition $b=b_{max}$ sets the maximum value of $\theta$, namely $\theta_{max}$.

Considering a polar cap of an angular radius of $\alpha$ centered at $\theta=\theta_{0}$, 
a function $h(\theta ; \alpha ,\theta_{0})$ is then defined as the range of $\phi$ included 
in the ``one-dimensional slice" at $\theta$ of the polar cap (cf. Fig.~6 in Hui \& Cheng 2004).
If $\theta_{0}+\alpha\leq\theta_{max}\leq 180^{o}$ and $\theta_{0}-\alpha\geq 0$, 
then $h(\theta ; \alpha ,\theta_{0})$ is defined as:
\begin{equation}
h(\theta ; \alpha ,\theta_{0})=\left\{\begin{array}{ll}
                               2\cos^{-1}\left(\frac{\cos\alpha -\cos\theta_{0}\cos\theta}{\sin\theta_{0}\sin\theta}\right) & \mbox{for $\theta_{0}-\alpha\leq\theta\leq\theta_{0}+\alpha$};\\
                               0 & \mbox{for $\theta$ outside the range $\theta_{0}\pm\alpha$}\end{array}\right.
\end{equation} 

For generating the light curves, $\theta_{0}$ is expressed as a function of time/rotational phase. 
Let $\beta$ is the angle between the axis of rotation of the star and the line joining 
the center of the polar cap and the center of the star, and $\gamma$ is the angle between 
the axis of rotation and the $z$-axis, then
\begin{equation}
\cos\theta_{0}=\sin\left(\beta\right)\sin\left(\gamma\right)\cos\left(\Omega t\right) + \cos\left(\beta\right)\cos\left(\gamma\right)
\end{equation}
where $\Omega$ is the rotational frequency of the star.
The relative brightness can be expressed as a function $\theta_{0}$, $M/R$ and 
$\alpha$:

\begin{equation}
A\left(\theta_{0};M/R,\alpha\right)=\left(1-\frac{2GM}{c^{2}R}\right)^{2}\left(\frac{GM}{c^{2}R}\right)^{2}\int_0^{x_{max}} h\left(x; \alpha ,\theta_{0}\right)x\,dx
\end{equation}
\noindent where $x=\frac{c^{2}b}{GM}$ and $x_{max}=\frac{c^{2}b_{max}}{GM}$.

With a view to minimize the number of free parameters in 
the modeling, we utilized the results from other analysis. 
From the best-fit BB model of pulsed spectrum, the polar cap size is found to be $\sim320$~m if \psr\ locates at a distance of 2~kpc. 
Assuming a neutron star radius of $R\sim10$~km, we fixed the angular radius of the polar cap at $\alpha\sim2^{\circ}$. 
From modeling the $\gamma$-ray light curve, Trepl et al. (2010)
suggest the viewing angle can possibly be in a range of 83-87$^{\circ}$. 
For a given mass-to-radius ratio, $\frac{GM}{c^{2}R}$, the effect of varying
the viewing angle in such small range in the pulse profile is negligible.
Also, for a polar cap with a small angular radius of $\sim2^{\circ}$,
it is likely that only one pole will cross the line-of-sight.  This scenario is
supported by the observed single broad peak. 
With these constraints, we 
minimized the number of free parameters by assuming a simple orthogonal rotator ($\gamma=\beta=90^{\circ}$)
with a single pole contribution. This leaves the
$\frac{GM}{c^{2}R}$ to be the only parameter for modeling the X-ray
light curve.
The best-fit model yields $\frac{GM}{c^{2}R}=0.21$ and a goodness-of-fit of $\chi^{2}=27.1$ for 31 D.O.F.. For 
$R\sim10$~km, it implies a neutron star mass of $M\sim1.4M_{\odot}$. The comparison of the best-fit model and the observed light curve 
is shown in Figure~\ref{pulse_sim}. For a conservative estimate, the $90\%$ confidence interval for 1 parameter of interest 
(i.e. $\Delta\chi^{2}=2.71$ above the minimum) is found to be $0.17<\frac{GM}{c^{2}R}<0.25$, which corresponds 
to $M\sim 1.2-1.7 M_{\odot}$ for $R\sim10$~km. Analysis with deeper follow-up observation can provide a tighter constraint on the 
mass-to-radius ratio of this neutron star. 

\subsection{Imaging spectroscopy of the central and southeastern regions of \G}
For investigating the X-ray emission from \G, we only focused on the extended features with relatively high surface brightness which 
are highlighted by the solid-line ellipses in Fig.~\ref{xmm_color} (referred as southeastern region hereafter) 
and Fig.~\ref{pn_img} (referred as central region hereafter).
Before the spectra of these extended structures were extracted, contributions from all the resolved point sources were firstly 
subtracted from the data. The background spectra for the southeastern region and the central region were sampled from 
the nearby low count regions as illustrated by the dash-ellipse in Fig~\ref{xmm_color} and dash-circle in Fig.~\ref{pn_img} respectively. 
The response files for this extended source analysis are generated
by \emph{rmfgen} and \emph{arfgen} with uniform spatial averaging. 
The spectra obtained from different cameras were binned dynamically so as to achieve a comparable signal-to-noise ratio.

By inspecting the X-ray spectrum of the central region (see Fig.~\ref{cocoon}), some emission line features such as 
Mg at $\sim1.4$~keV and Si at $\sim1.9$~keV can be clearly seen. 
This prompts us to examine the spectrum with an absorbed collision ionization equilibrium (CIE) plasma model (XSPEC model: VEQUIL). 
To examine whether the metal abundance of \G\ deviates from the solar values, we thawed the corresponding parameters individually to see if
the goodness of fit can be improved.
The best-fit model yields a plasma temperature of $kT\sim0.6$~keV and a column absorption of $N_{H}\sim10^{22}$~cm$^{-2}$. 
There is also indication that Mg is overabundant in comparison with the solar values. However,
such model cannot provide an adequate description of the observed data even with metal abundance Mg open as free 
parameter ($\chi^{2}=391.53$ for 159 D.O.F.). By examining the fitting residuals, systematic deviations at energies 
beyond $\sim2$~keV are noted. We suspected that an extra PL component might be required. With the additional PL component, 
the goodness-of-fit has been significantly improved ($\chi^{2}=209.36$ for 157 D.O.F.). It yields a
column absorption of $N_{H}=8.2^{+0.8}_{-1.0}\times10^{21}$~cm$^{-2}$, a plasma temperature of 
$kT=0.59^{+0.02}_{-0.03}$~keV, a Mg abundance of $1.9^{+0.4}_{-0.3}$ with respect to the solar value,
a emission measure of $\int_{V}n_{e}n_{H}dV=(2.8\pm0.9)\times10^{11}D^{2}$~cm$^{-3}$
and a PL index of $\Gamma=1.6^{+0.5}_{-0.2}$, where $n_{e}$, $n_{H}$, $V$ and $D$ are the electron density (cm$^{-3}$), hydrogen density 
(cm$^{-3}$), volume of interest (cm$^{3}$) and the source distance (cm). 

For further improving the spectral modeling, we examined the fitting residuals of the CIE+PL fit. We noticed there are scattering 
of the residuals at energies greater than $\sim2$~keV which are probably stemmed from the additonal PL. As demonstrated by 
Huang et al. (2014), the residuals in the hard band can possibly resulted from the residual soft proton contamination
in individual cameras after the data screening. And therefore, instead of originating from the particle acceleration, 
the additional PL component merely provides a phenomenological description for 
such residual soft proton background with the PL index and normalization vary among different EPIC cameras. By disentangling the PL 
component in MOS1, MOS2 and PN, we found the goodness-of-fit can be further improved ($\chi^{2}=175.04$ for 153 D.O.F.). In view of the 
different best-fit PL index inferred from different cameras, we conclude that the residuals in the hard band are contributed by the residual
background. Under this consideration, the best-fit absorbed CIE component yields $N_{H}=7.9^{+0.8}_{-0.5}\times10^{21}$~cm$^{-2}$, 
$kT=0.60^{+0.02}_{-0.03}$~keV, a Mg abundance of $2.0\pm0.4$ with respect to the solar value and 
$\int_{V}n_{e}n_{H}dV=2.5^{+0.8}_{-0.5}\times10^{11}D^{2}$~cm$^{-3}$. 

We have also examined the central part of \G\ with an non-equilibrium ionization (NEI) model (XSPEC model: VNEI). With 
additional PL component applied to account for the residual soft proton contamination in the individual camera, we found that 
the NEI model results in a further improved goodness-of-fit ($\chi^{2}=167.95$ for 152 D.O.F.). It yields
a column absorption of $N_{H}=7.4^{+1.0}_{-1.4}\times10^{21}$~cm$^{-2}$, a plasma temperature of
$kT=1.6^{+0.6}_{-0.3}$~keV, a Mg abundance of $1.6\pm0.2$ with respect to the solar value,
a emission measure of $\int_{V}n_{e}n_{H}dV=7.4^{+1.0}_{-0.8}\times10^{10}D^{2}$~cm$^{-3}$
and an ionization timescale of $n_{e}t=1.8^{+0.6}_{-0.4}\times10^{10}$~s~cm$^{-3}$ where $n_{e}$ and $t$ are 
electron density and time elapsed since the gas has been shock-heated respectively. We noted that the inferred plasma temperature
is significantly higher than the CIE fit. This can be due to the effect that the ionization states for a plasma in NEI at 
a given temperature are lower than those in the CIE situation (e.g. see Fig.~11 in Vink 2012). Assuming a CIE condition can thus result in an 
underestimation of the plasma temperature. The NEI condition is further indicated by the best-fitted ionization timescale which is 
significantly less than that ($n_{e}t\sim10^{12}$~s~cm$^{-3}$) required to reach CIE (cf. Vink 2012). In view of these, 
the NEI scenario is preferred. 

For fitting the spectrum in the southeastern region, we have also considered both CIE and NEI models. Same as the situation in the analysis 
of the diffuse X-rays from the central region, 
additional PL component were included for modeling the residual soft proton contamination in MOS1 and MOS2 individually. 
The CIE yields $N_{H}=4.4^{+1.0}_{-0.7}\times10^{21}$~cm$^{-2}$, 
$kT=0.63^{+0.02}_{-0.03}$~keV, a Mg abundance of $3.3\pm0.8$ with respect to the solar value and 
$\int_{V}n_{e}n_{H}dV=5.1^{+2.5}_{-1.1}\times10^{11}D^{2}$~cm$^{-3}$ ($\chi^{2}=305.11$ for 242 D.O.F.). 
In comparison, the NEI model results in an improved goodness-of-fit ($\chi^{2}=261.06$ for 241 D.O.F.). 
It yields $N_{H}=(7.4\pm0.6)\times10^{21}$~cm$^{-2}$,
$kT=1.1^{+0.5}_{-0.3}$~keV, a Mg abundance of $1.2\pm0.1$ with respect to the solar value, $\tau=n_{e}t=2.4^{+1.9}_{-0.7}\times10^{10}$~s~cm$^{-3}$ 
and $\int_{V}n_{e}n_{H}dV=8.2^{+3.4}_{-1.9}\times10^{11}D^{2}$~cm$^{-3}$. Both the goodness-of-fit and the small value of $n_{e}t$ resulted from 
this fit suggest the remnant emission in this region is also in an NEI condition.

\section{DISCUSSION}
We have reported the results from a detailed X-ray analysis of \psr\ and \G. The column absorption deduced from the X-ray spectra of \psr\ 
(see \S2.3) is consistent with that deduced from various parts of the diffuse emission (\S2.4). 
And we note that it is also consistent with neutral hydrogen density inferred from the HI absorption spectrum (Leahy et al. 2013). These 
indicate that the pulsar emission, diffuse X-ray emission and the radio shell are essentially at the same distance. Hence, the association 
between \psr\ and \G\ is supported by our investigation. 

\begin{figure*}[b]
\centerline{\psfig{figure=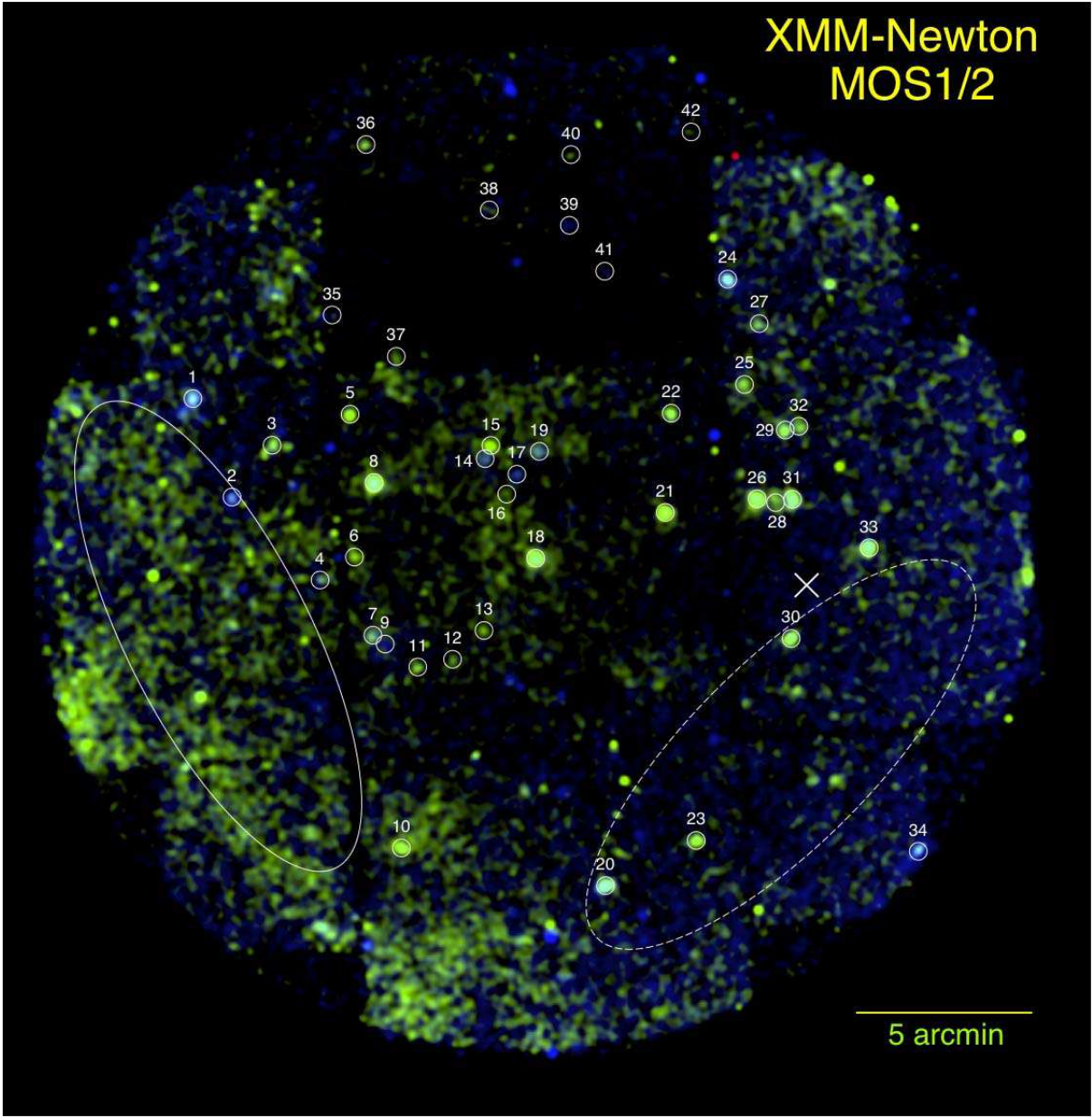, width=16cm,clip=,angle=0}}
\caption[]{Vignetting-corrected {\it XMM-Newton} MOS1/2 color image of the field around \psr\ 
(red: 0.5-1~keV, green: 1-2~keV, blue: 2-10~keV). The binning factor of this image is 1". 
Adaptive smoothing has been applied to achieve a minimum signal-to-noise ratio of 3. The white 
cross illustrates the the nominal 
geometrical center of SNR \G\ (Green 2009). Soft diffuse emission is found
in the field. The overlaid solid-line ellipse illustrates the extraction region for
the diffuse spectra of the southeastern rim of \G. The dashed ellipse shows the background region used in the remnant analysis. 
Including \psr\ (source 18), 42 X-ray point sources 
detected by this observation are highlighted with the labels consistent with Table~1. 
Top is north and left is east.}
\label{xmm_color}
\end{figure*}

Given the pulsar-SNR association and assuming the birth place of \psr\ is not far away from the geometrical center of \G, we estimated 
the projected velocity of the pulsar. The angular separation between \psr\ and the geometrical center is $\sim0.1^{\circ}$ (cf. Fig.~\ref{xmm_color}).
At a distance of 2~kpc, this corresponds to a physical separation of $1.4\times10^{14}$~km. Together with a Sedov age of $\sim8000$~yrs deduced 
for \G\ (Leahy et al. 2013), the magnitude of the projected velocity of \psr\ is expected to be $v_{p}\sim550$~km~s$^{-1}$ which is not unreasonable 
for the known pulsar population (Hobbs et al. 2005). The projected direction of the pulsar motion is indicated by the arrow in Fig.~\ref{cxc_pwn_img}.
The speculated pulsar motion can possibly be checked by a dedicated $\gamma-$ray timing analysis of the full time-span \emph{Fermi} LAT data 
in further studies. 

Such speculated pulsar velocity should be far exceeding the local speed of sound. For a pulsar moving supersonically, it is expected to 
drive a bow shock through the ambient medium. The pulsar wind particles will be accelerated and produce synchrotron X-ray emission. With 
the motion of pulsar, this will result in a cometary-like nebula as the extended structure found in the high resolution {\it Chandra} image 
(see Fig.~\ref{cxc_pwn_img} and Fig.~\ref{brightness}). 
In this case, the termination shock radius $R_{s}$ is determined by the 
ram pressure balance between the relativistic pulsar wind particles and the circumstellar medium at the head of the shock: 

\begin{equation}
R_{s}\simeq\left(\frac{\dot{E}}{2\pi\rho v_{p}^{2}c}\right)^{1/2}
\sim 3\times10^{16}\dot{E}_{34}^{1/2}n^{-1/2}v_{p,100}^{-1} {\rm cm},
\end{equation}

\noindent where $v_{p,100}$ is the velocity of the pulsar in units of 100 km s$^{-1}$,
$\dot{E}_{34}$ is the spin-down luminosity of the pulsar in units of $10^{34}$
erg s$^{-1}$, and $n$ is the number density of the circumstellar medium in units of cm$^{-3}$.

\begin{figure*}[b]
\centerline{\psfig{figure=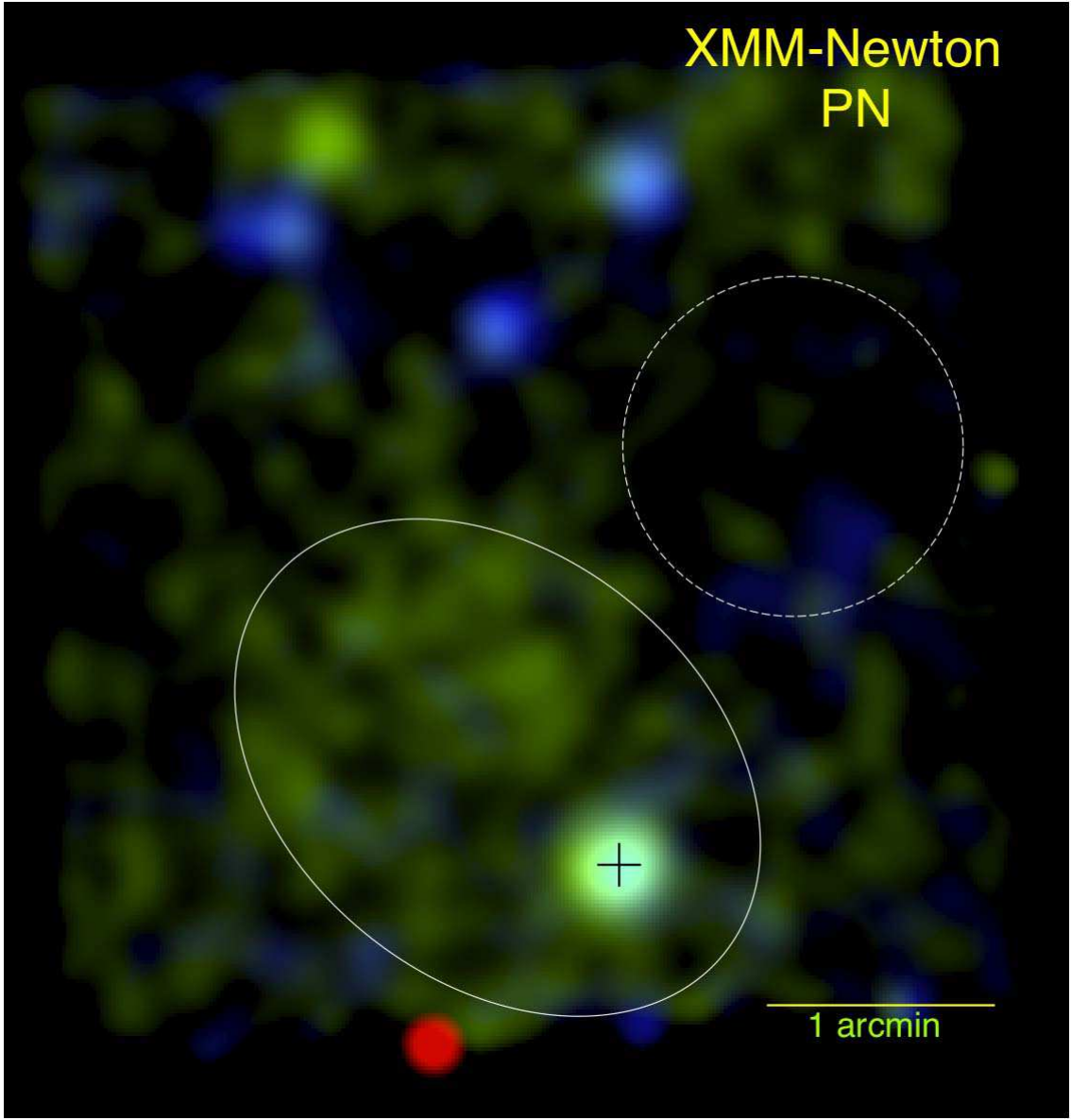, width=15cm,clip=,angle=0}}
\caption[]{ The sky region around
\psr\ (illustrated by the black cross) as seen by {\it XMM-Newton} PN camera in small-window mode. This image is binned, color-coded,
vignetting-corrected and
adaptively smoothed in the same way as Fig.~\ref{xmm_color}. Soft emission around the pulsar can be clearly seen.
A cocoon-like feature around \psr\ is highlighted by a solid-line ellipse which is adopted as the extraction region for 
the spectra from all EPIC cameras. The dashed circle shows the background region used in the remnant analysis.
The bright red spot at the bottom is an artifact at edge of the window.
Top is north and left is east.}
\label{pn_img}
\end{figure*}

\begin{figure*}[b]
\centerline{\psfig{figure=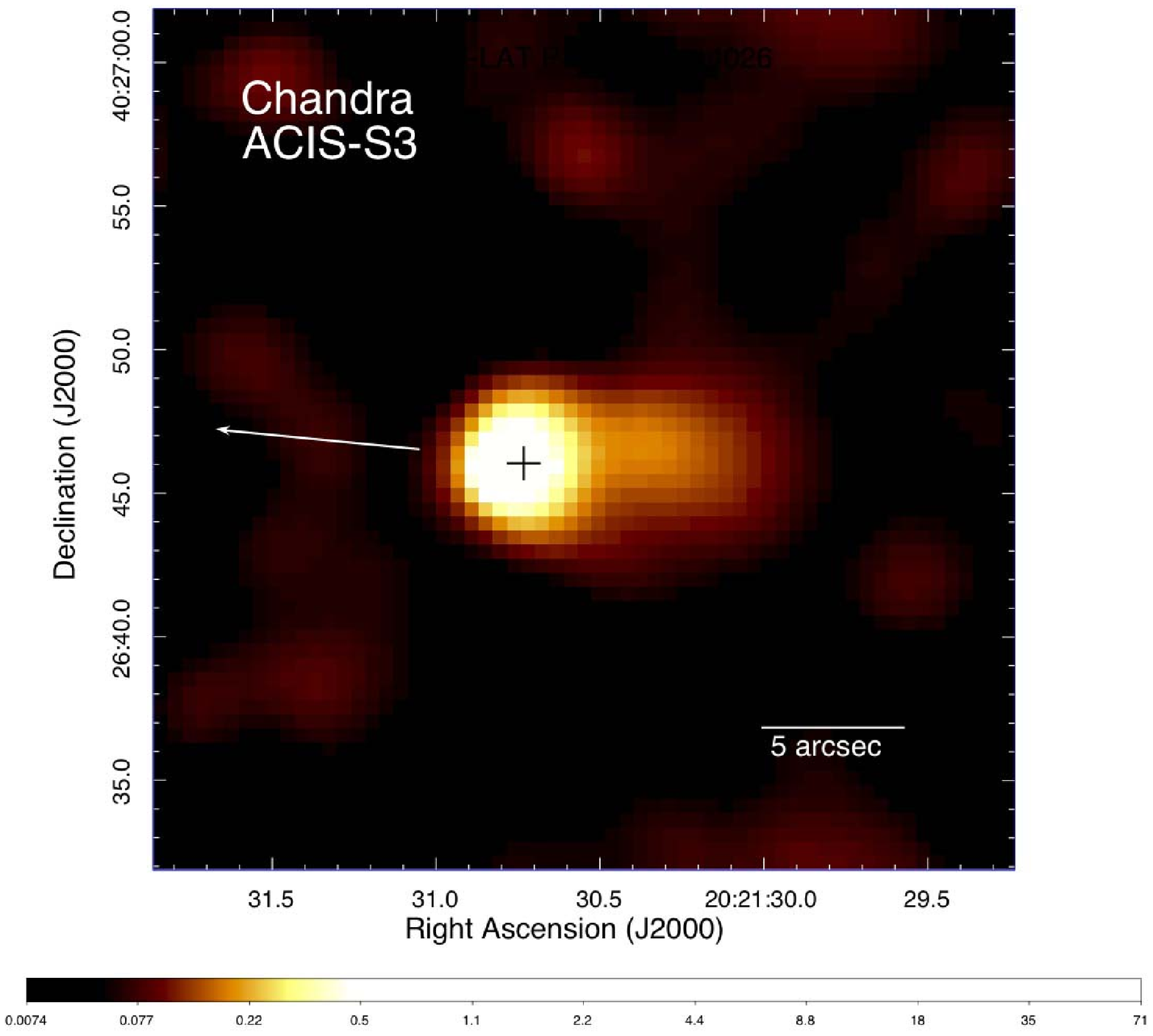, width=16cm,clip=,angle=0}}
\caption[]{The $30"\times30"$ X-ray image in the energy band 0.3-8~keV around \psr\ as seen 
by {\it Chandra} ACIS-S3 CCD. The 
X-ray position as determined by Weisskopf et al. (2011) is illustrated by the black cross. The binning 
factor of the image is $0.5"$ and has been adaptively smoothed to achieve a minimum signal-to-noise ratio 
of 3. A nebular structure extends to the west from the pulsar can be clearly seen. The white arrow 
illustrates the speculated proper motion direction by extrapolating the nominal center of \G\ to the current pulsar position.}
\label{cxc_pwn_img}
\end{figure*}

\begin{figure*}[b]
\centerline{\psfig{figure=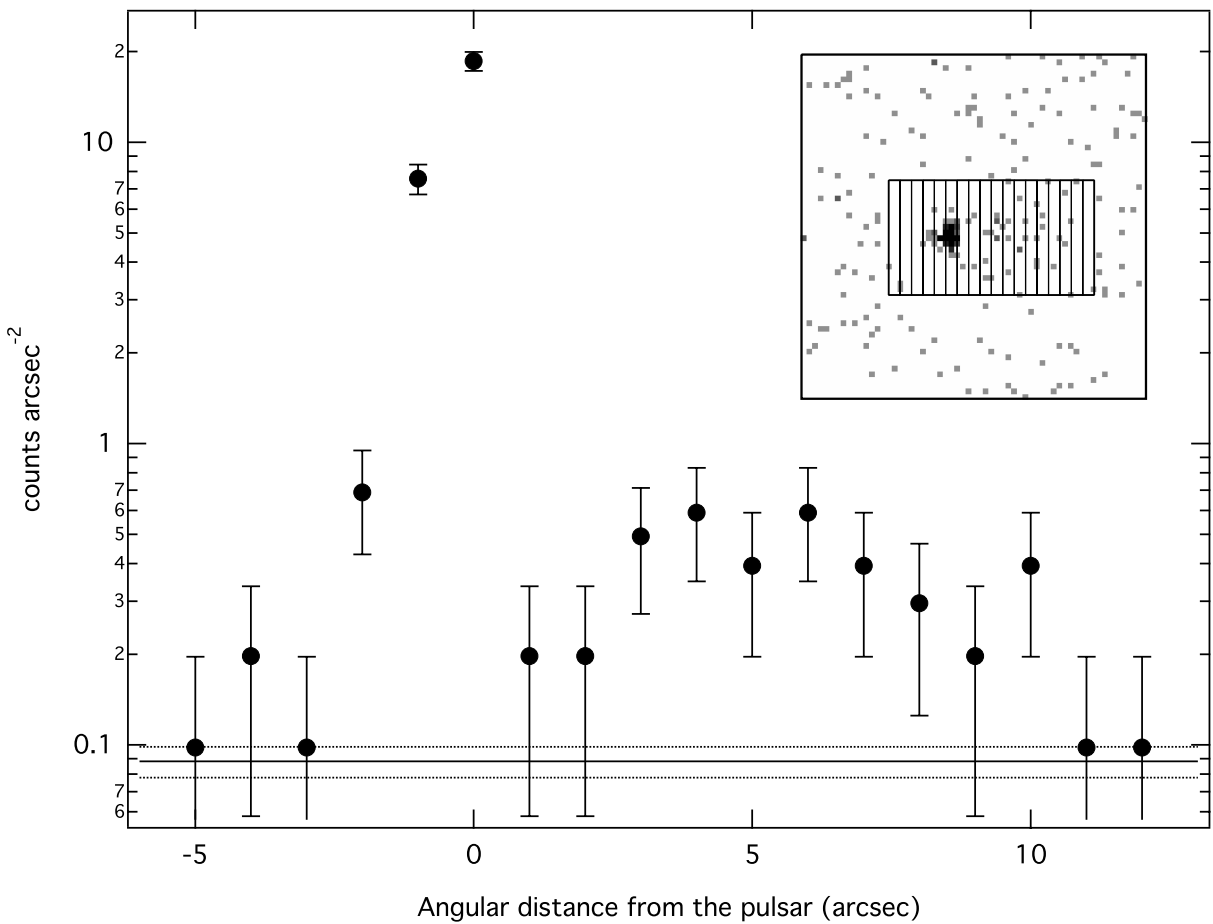, width=16cm,clip=,angle=0}}
\caption[]{The X-ray brightness profile in the energy band of $0.3-8$~keV along the orientation 
of the PWN associated with \psr\ as observed by {\it Chandra} ACIS-S3 CCD (cf. Fig.~\ref{cxc_pwn_img}).
The insets show the bins used in
computing the profile. Each bin has a size of $1''\times10''$. The average
background level and its $1\sigma$ deviation are indicated by horizontal lines which were calculated by sampling
from the source-free regions within a $2'\times2'$ field around the pulsar.}
\label{brightness}
\end{figure*}

\begin{figure*}[b]
\centerline{\psfig{figure=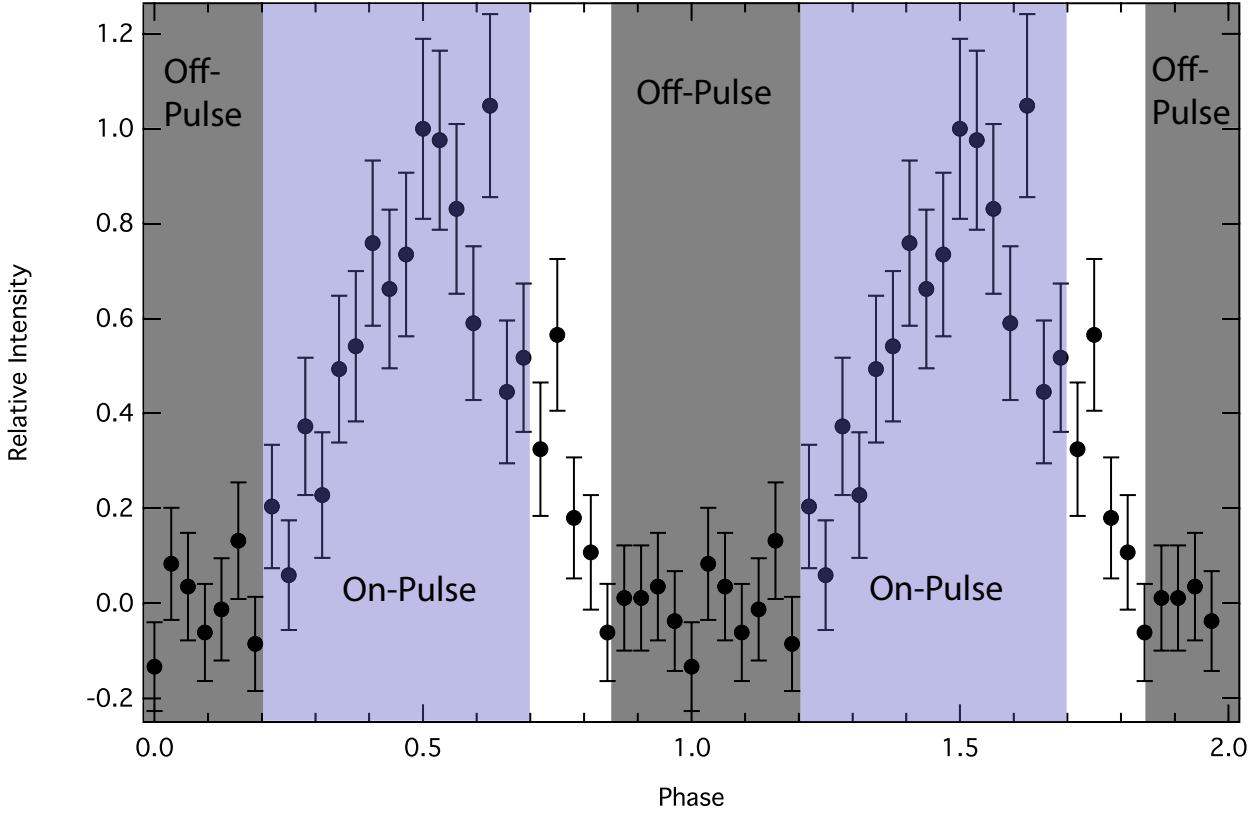, width=17cm,clip=}}
\caption[]{The pulsation and quiescent stage labelled by the pulse profile of PSR~J2021+4026 (0.7-2~keV) as observed 
by {\it XMM-Newton}/PN camera. The obtained light curve was folded with a spin frequency of 3.768995206~Hz. 
Two rotation cycles are shown for clarity. 
Error bars indicate the $1\sigma$ uncertainty. 
The grey shaded regions and the blue shaded regions illustrate the off-pulse phase (i.e. DC level) and the on-pulse phase 
respectively. The pulsed spectrum was obtained by subtracting the DC level from the source spectrum extracted from the on-pulse 
phase interval.}
\label{pulse_phase}
\end{figure*}

\begin{figure*}[b]
\centerline{\psfig{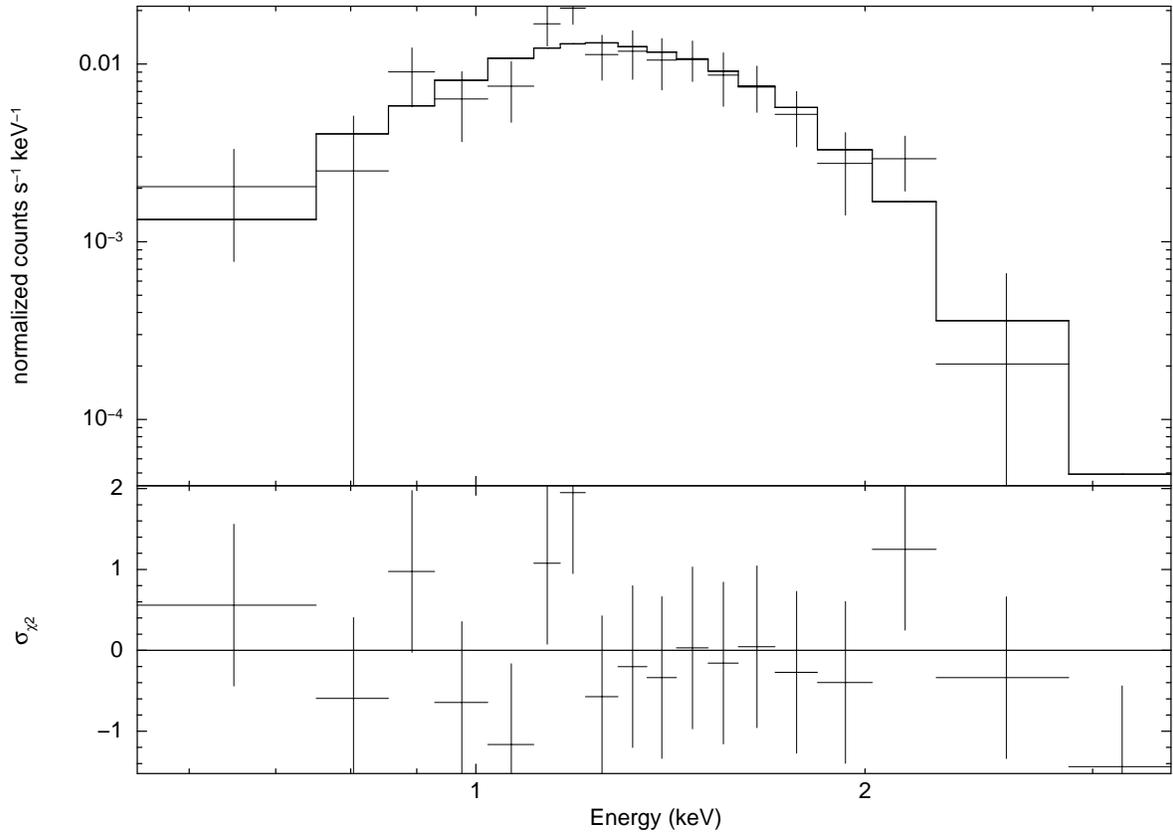}}
\caption[]{The pulsed spectrum of \psr\ as observed by {\it XMM-Newton} PN camera 
with the best-fit blackbody model illustrated (upper panel)
and the contributions to the $\chi^{2}$ fit statistic
(lower panel).}
\label{psr_spec}
\end{figure*}

For constraining $n$, we utilized the results inferred by the NEI plasma model fit of the central extended X-ray emission of \G\ which surrounds 
\psr\ (see Fig.~\ref{pn_img}). The best-fit emission measure of this feature allows us to estimate the hydrogen density $n_{\rm H}$ 
and the electron density $n_{\rm e}$ in this circumstellar region. Assuming $n_{e}$ and $n_{H}$ 
are uniform in the extraction region and distance of 2~kpc, the emission measure can be approximated by 
$n_{e}n_{H}V\sim2.8\times10^{54}$~cm$^{-3}$. We further assumed a geometry of oblated spheroid for the spectral extraction region,
the volume of interest is $V\sim4\times10^{55}$~cm$^{3}$. For a fully ionized plasma with
$\sim10\%$ He ($n_{\rm e}\sim1.2n_{\rm H}$), $n_{H}$ is estimated as $\sim0.24$~cm$^{-3}$. Together with the spin-down power of 
$\dot{E}_{34}=10$ and our speculated $v_{p,100}\sim5.5$, Equation~5 implies a termination radius of $R_{s}\sim3.5\times10^{16}$~cm. 
It corresponds to a stand-off angle of $\sim1"$ ahead of the pulsar at a distance of 2~kpc. Comparing this estimate to the
angular size of the cometary-like feature behind the pulsar (see Fig.~\ref{cxc_pwn_img}), 
the ratio of termination shock radii between the directions immediately behind and directly ahead of the pulsar is estimated to be $\sim10$
which is comparable with the ratios observed in other fast-moving pulsar such as PSR~J1747-2958 (Gaensler et al. 2004). 

At 2~kpc, the physical size of the synchrotron X-ray nebula associated with \psr\ is $l_{\rm pwn}\sim3\times10^{17}$~cm. This implies the timescale 
for the pulsar to traverse its nebula is $\tau_{\rm pwn}=l_{\rm pwn}/v_{p}\sim170$~yrs. The magnetic field strength of the nebula can be estimated 
by assuming $\tau_{\rm pwn}$ is comparable with the synchrotron cooling timescale of the electrons:

\begin{equation}
\tau_{\rm syn}=\frac{6\pi m_{e}c}{\gamma\sigma_{T}B^{2}}\simeq 10^{5}\left(\frac{h\nu_{X}}{\rm keV}\right)^{-\frac{1}{2}}B_{\mu G}^{-\frac{3}{2}} {\rm yrs}
\end{equation}

\noindent where $\gamma$ is the Lorentz factor of the wind, 
$\sigma_{T}$ is the Thompson cross
section, $\nu_{X}=3\gamma^{2}eB/2m_{e}c$ is the characteristic synchrotron frequency and $B_{\mu G}$ is the
magnetic field in the shocked region in unit of micro gauss. This suggests the nebular magnetic strength is at the order of $\sim15$~$\mu G$ for 
a characteristic energy of $h\nu_{X}\sim$1~keV. 

The magnetic field estimate can further enable us to compute the electron synchrotron
cooling frequency $\nu_{\rm c}$:

\begin{equation}
\nu_{\rm c}=\frac{18\pi em_{e}c}{\sigma_{T}^{2}\tau^{2}_{\rm syn}B^{3}}
\end{equation}

\noindent which is estimated to be $\sim1.7\times10^{19}$~Hz (i.e. $h\nu_{\rm c}\sim70$~keV). Since this is far exceeding the observed frequencies, 
it suggests the X-ray emission of the nebula is in a slow cooling regime (Chevalier 2000; Cheng, Taam \& Wang 2004). 
In this regime, electrons with the energy distribution,
$N\left(\gamma\right)\propto\gamma^{-p}$, are able to radiate their energy
in the trail with photon index $\Gamma=(p+1)/2$. The index $p$ due to shock
acceleration typically lies between 2 and 3 (cf. Cheng et al. 2004 and reference
therein). This would result in a photon index $\sim1.5-2.0$. This is consistent with the observed photon index of $\Gamma\sim1.5$ for the 
unpulsed non-thermal spectral component of \psr.

\begin{figure*}[b]
\centerline{\psfig{figure=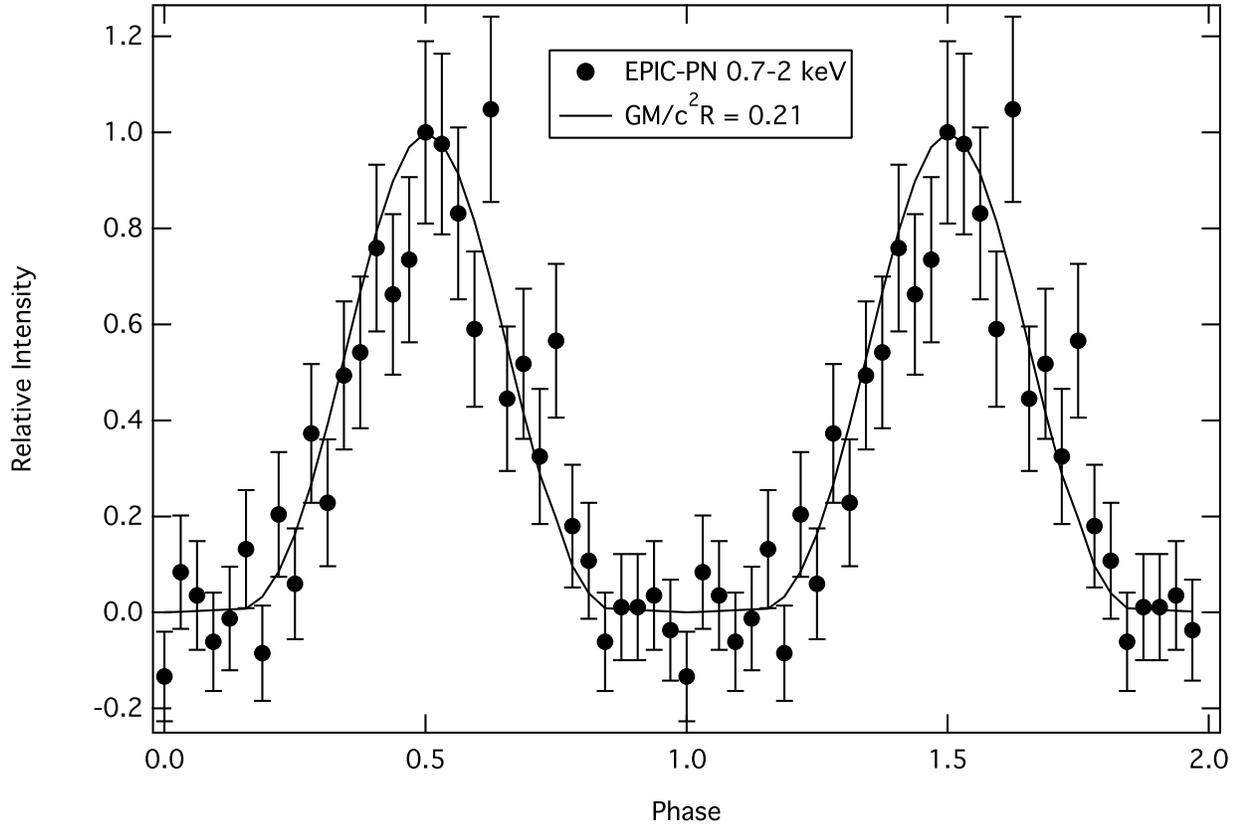, width=17cm,clip=}}
\caption[]{The pulse profile of \psr\ as observed by {\it XMM-Newton} PN camera in 0.7-2.0~keV (cf. Fig.~1 in Lin et al. 2013) 
and the best-fit simulated profile (solid curve) with the effects of gravitational light-bending incoporated. 
Error bars indicate the $1\sigma$ uncertainty. Two rotation cycles are shown for clarity.}
\label{pulse_sim}
\end{figure*}

\begin{figure*}[b]
\centerline{\psfig{figure=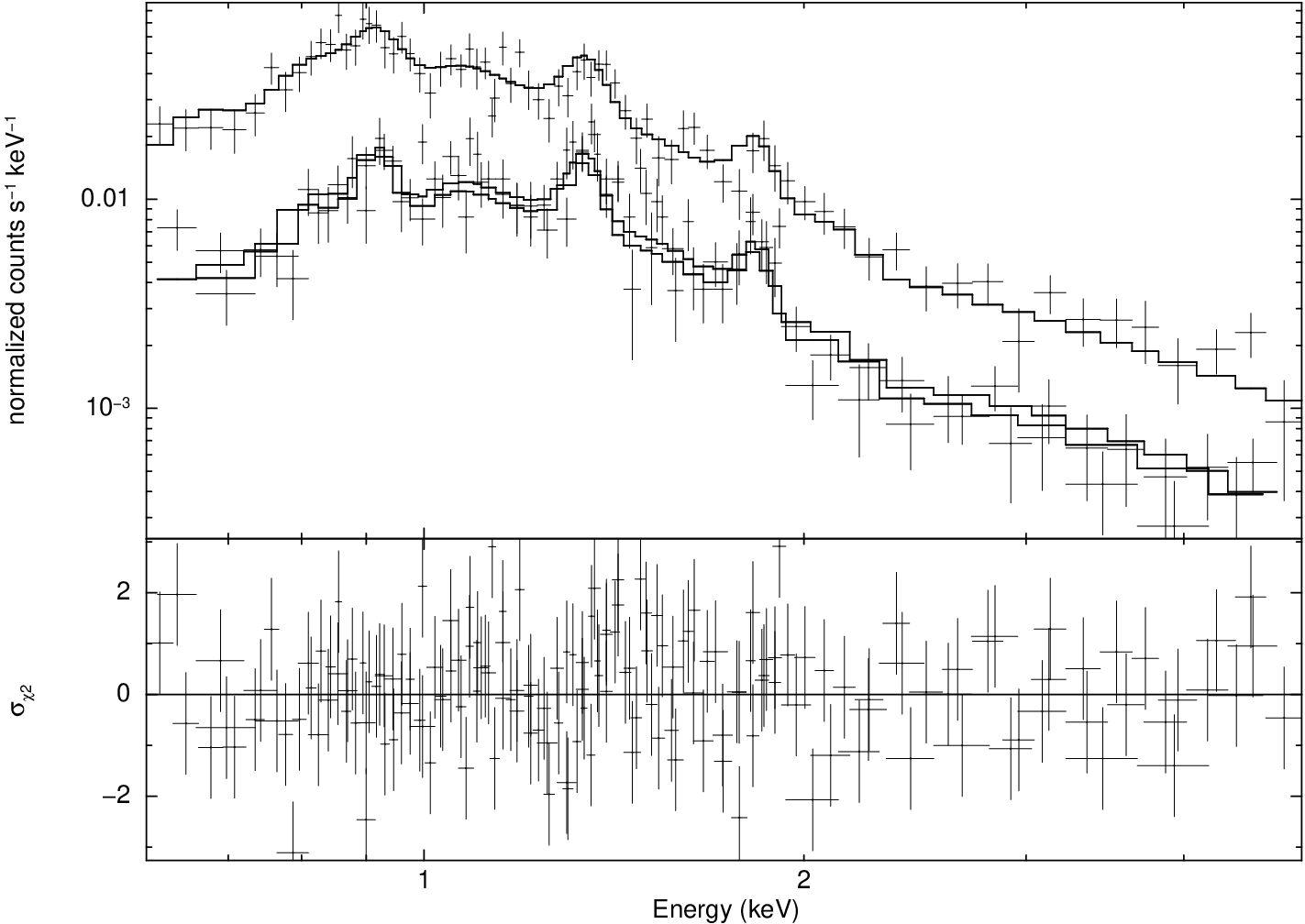, width=11cm,clip=,angle=-90}}
\caption[]{Upper panel: X-ray energy spectra of the central region of \G\ as observed by MOS1/2 (cf. Fig.~\ref{xmm_color}) and 
PN (cf. Fig.~\ref{pn_img}) which are simultaneously fitted to an absorbed non-equilibrium ionization plasma 
model. Additional power-law components have been applied to account for the residual soft proton contamination in the 
individual camera. Lower panel: contributions to the $\chi^{2}$ fit statistic.} 
\label{cocoon}
\end{figure*}

For the X-ray pulsation of \psr, our analysis of the pulsed spectrum confirmed its thermal origin. We noted that the best-fit blackbody radius
is comparable with the polar cap size of $\sqrt{\frac{2\pi R}{cP}}\sim300$~m by adopting a dipolar field geometry, a neutron star radius of 
$R=$10~km and a rotational period of $P=265$~ms. This suggests the thermal emission is originated from the hot polar cap. Such assertion 
is further demonstrated by the agreement between the X-ray pulse profile and the simulated modulation by a hot spot on the surface of a 
canonical rotating neutron star (Fig.~\ref{pulse_sim}). 

The most remarkable properties of \psr\ are the sudden changes of its spin-down rate, pulse profile and flux in $\gamma-$ray on a 
timescale shorter than a week (Allafort et al. 2013). The authors speculated that such abrupt changes are resulted from a 
shift in the magnetic field structure which in turn leads to the change of either magnetic inclination and/or the effective current 
(see discussion in Allafort et al. 2013). These can be precipitated by a reconfiguration of the magnetic field line footprints on the stellar 
surface. The polar cap size is defined by the footprint of the last open-field lines and its temperature is determined by the backflow current 
from the accelerating region. Therefore, according to the scenario proposed by Allafort et al. (2013), one should expect a correlated change 
in the thermal X-ray flux and/or the X-ray pulse profile. 

We have attempted to look for such expected X-ray change across the 
$\gamma-$ray jump by a joint analysis of \emph{Chandra} and \emph{XMM-Newton} data.
Although we did not find any conclusive variability, we would like 
to point out that the significance of the analysis is limited by the small photon statistic of \emph{Chandra} data. 
Also, the poor temporal 
resolution of this \emph{Chandra} data does not allow any investigation of the X-ray pulsation. For a follow-up investigation of this unique
pulsar, we encourage a long-term coordinated X-ray and $\gamma-$ray monitoring with \emph{XMM-Newton} and \emph{Fermi} which can provide a 
better understanding the nature of its variability. 

Apart from \psr, we have examined the diffuse X-ray emission of \G\ in the FoV of our \emph{XMM-Newton} observation (See \S2.4).
Leahy et al. (2013) have also analysed the central region with \emph{Chandra} data and obtained 
$N_{H}=(7.5-11.1)\times10^{21}$~cm$^{-2}$, $kT=0.6-2.7$~keV and $n_{e}t=(1.7-12)\times10^{10}$~s~cm$^{-3}$. 
Within the tolerence of $1\sigma$ uncertainties, our results are consistent with theirs. With the much 
improved photon statistic of our \emph{XMM-Newton} data, we constrained the spectral parameters $N_{H}$, $kT$ and 
$n_{e}t$ to an accuracy of $\sim32\%$, $\sim56\%$ and $\sim56\%$ respectively. 
For comparison, the uncertainties of the corresponding parameters reported by Leahy et al. (2013) are 
$\sim39\%$, $\sim203\%$ and $\sim312\%$. 

For the southeastern rim, Uchiyama et al. (2002) have investigated its X-ray properties with \emph{ASCA} (R2 region in their work). 
They have modeled the spectrum with a CIE model and obtained a temperature $kT=0.53\pm0.07$~keV which is consistent with our CIE estimate.
However, the quality of \emph{ASCA} data did not allow the authors to discern whether the X-ray emission is in CIE or NEI state. 

In our study, we confirmed that the remnant 
emission from our investigated regions are in NEI state. This is probably due to the low electron density and the time elapsed since the gas has 
been shock-heated is not long enough for the plasma to reach the equilibrium. The best-fit emission measures and the ionization timescales 
allow us to estimate these quantities. From the above discussion, the electron density of the central region is found to be 
$n_{e}\sim0.3$~cm$^{-3}$. With the best-fit ionization timescale of $n_{e}t\sim1.8\times10^{10}$~s~cm$^{-3}$, the elapsed time since 
the arrival of the shock is estimated as $t\sim1900$~yrs. For the diffuse emission in the southeastern region, the emission measure and the 
volume of interest are $n_{e}n_{H}V\sim3.1\times10^{55}$~cm$^{-3}$ and $V\sim10^{57}$~cm$^{-3}$ respectively. This implies a electron 
density of $n_{e}\sim0.2$~cm$^{-3}$ in this region. Together with the ionization timescale of $n_{e}t\sim2.4\times10^{10}$~s~cm$^{-3}$ inferred 
for this region, this suggests the gas has been shock-heated $\sim3800$~yrs ago. Such age estimates are significantly smaller than the Sedov age 
of \G\ (Leahy et al. 2013). This might indicate that these plasma have been heated by the reverse shock(s) that have returned to the remnant center 
not long ago. We would like to point out this interpretation is stemmed from the spectral fitting with a relatively simple NEI model. 
For example, the temperatures of different plasma constituents for such small ionization timescale are not neccessary to equilibrate. A more 
sophisticated modeling of the observed remnant spectrum around the center can help to confirm if the reverse shocks have arrived yet. 

\acknowledgments{
CYH is supported by the Chungnam National University research fund in 2014.
CPH and YC are supported by the Ministry of Science and Technology through the grant NSC 102-2112-M-008-020-MY3.
JT and KSC are supported by a 2014 GRF grant of Hong Kong Government under HKU 17300814P.
AKHK is supported by the Ministry of Science and Technology of Taiwan
through grants 100-2628-M-007-002-MY3, 100-2923-M-007-001-MY3 and 103-2628-M-007-003-MY3.
}

\begin{figure*}[b]
\centerline{\psfig{figure=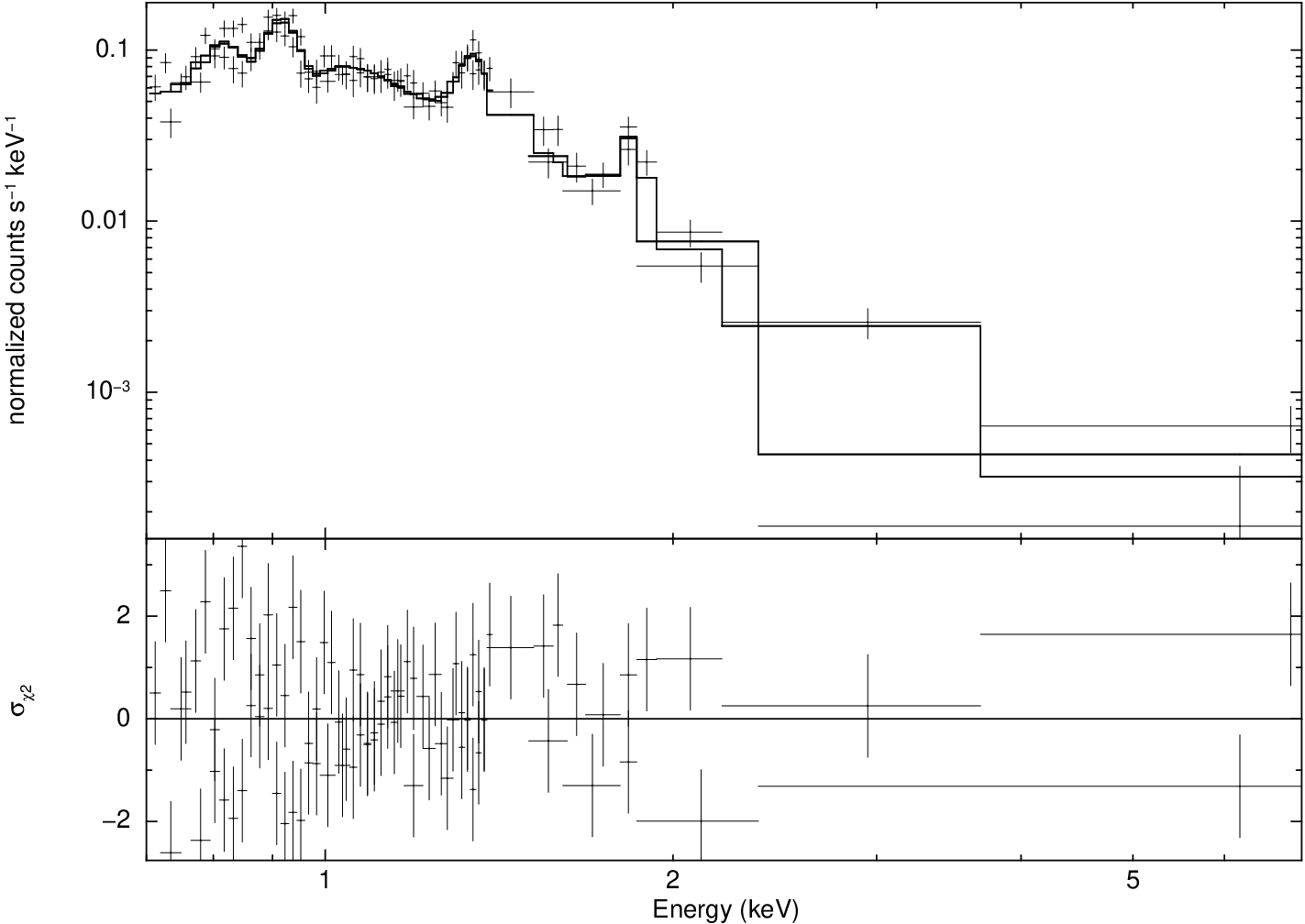, width=11cm,clip=,angle=-90}}
\caption[]{Upper panel: X-ray energy spectrum of southeastern region of \G\ as observed by MOS1/2 (cf. Fig.~\ref{xmm_color}) 
which are simultaneously fitted to an absorbed non-equilibrium ionization plasma       
model. Additional power-law components have been applied to account for the residual soft proton contamination in the  
individual camera. Lower panel: contributions to the $\chi^{2}$ fit statistic.}
\label{dr4}
\end{figure*}

\begin{landscape}[b]
\begin{longtable}{lllccccc}
\caption{X-ray sources detected in the southeastern field of \G\ as labeled in Figure~\ref{xmm_color}}\\
 \hline\hline
Source$^{\rm (a)}$ & RA (J2000) & DEC (J2000) & $r_{1\sigma}^{\rm (b)}$ & \multicolumn{3}{c}{Net count rate}  & Remark$^{\rm (c)}$\\
        &            &             &               & MOS1  & MOS2 & PN & \\ 
No.	  &	(h:m:s)	 &	(d:m:s)	& (arcsec) &   (cts ks$^{-1}$) & (cts ks$^{-1}$) & (cts ks$^{-1}$)  &    \\
\hline
1 & 20:22:21.77 & 40:31:17.93 & 0.43 & 7.11 $\pm$ 0.57 & 7.27 $\pm$ 0.60 &  -  & - \\
2 & 20:22:15.89 & 40:28:30.17 & 0.44 & 3.01 $\pm$ 0.46 & 3.45 $\pm$ 0.39 &  -  & - \\
3 & 20:22:09.89 & 40:29:58.75 & 0.67 & 1.06 $\pm$ 0.28 & 1.64 $\pm$ 0.28 &  -  & - \\
4 & 20:22:02.69 & 40:26:08.79 & 0.47 & 1.17 $\pm$ 0.21 & 1.25 $\pm$ 0.30 &  -  & - \\
5 & 20:21:58.23 & 40:30:51.61 & 0.48 & 2.93 $\pm$ 0.31 & 2.51 $\pm$ 0.29 &  -  & - \\
6 & 20:21:57.57 & 40:26:48.07 & 0.52 & 1.25 $\pm$ 0.20 & 1.48 $\pm$ 0.23 &  -  & W44 \\
7 & 20:21:54.78 & 40:24:34.84 & 0.69 & 1.34 $\pm$ 0.22 & 1.38 $\pm$ 0.24 &  -  & - \\
8 & 20:21:54.65 & 40:28:55.28 & 0.21 & 11.69 $\pm$ 0.49 & 9.83 $\pm$ 0.49 &  -  & - \\
9 & 20:21:52.95 & 40:24:19.73 & 0.78 & 0.69 $\pm$ 0.19 & 0.77 $\pm$ 0.18 &  -  & - \\
10 & 20:21:50.50 & 40:18:32.05 & 0.55 & 3.62 $\pm$ 0.38 & 5.37 $\pm$ 0.50 &  -  & - \\
11 & 20:21:48.07 & 40:23:40.65 & 0.77 & 0.94 $\pm$ 0.21 & 1.10 $\pm$ 0.22 &  -  & - \\
12 & 20:21:42.89 & 40:23:54.06 & 0.72 & 0.75 $\pm$ 0.17 & 0.66 $\pm$ 0.18 &  -  & W38 \\
13 & 20:21:38.20 & 40:24:43.18 & 0.60 & 1.40 $\pm$ 0.19 & 1.45 $\pm$ 0.22 &  -  & W34 \\
14 & 20:21:38.05 & 40:29:36.34 & 0.30 & 0.88 $\pm$ 0.18 & 1.30 $\pm$ 0.22 & 3.30 $\pm$ 0.42 & W33 \\
15 & 20:21:37.18 & 40:29:58.72 & 0.32 & 2.53 $\pm$ 0.26 & 2.50 $\pm$ 0.27 & 7.65 $\pm$ 0.52 & W32 \\
16 & 20:21:34.81 & 40:28:35.74 & 1.23 & 0.00 $\pm$ 0.00 & 0.00 $\pm$ 0.00 & 1.34 $\pm$ 0.32 & W30 \\
17 & 20:21:33.30 & 40:29:09.82 & 0.47 & 0.86 $\pm$ 0.18 & 1.10 $\pm$ 0.25 & 3.21 $\pm$ 0.39 & W27 \\
18 & 20:21:30.48 & 40:26:46.30 & 0.22 & 4.79 $\pm$ 0.32 & 4.45 $\pm$ 0.33 & 12.90 $\pm$ 0.64 & W20 \\
19 & 20:21:29.95 & 40:29:48.68 & 0.38 & 1.70 $\pm$ 0.23 & 1.52 $\pm$ 0.23 & 4.89 $\pm$ 0.46 & W19 \\
20 & 20:21:20.07 & 40:17:27.78 & 0.46 & 5.52 $\pm$ 0.50 & 5.68 $\pm$ 0.55 &  -  & - \\
21 & 20:21:11.10 & 40:28:04.43 & 0.21 & 9.13 $\pm$ 0.44 & 7.82 $\pm$ 0.43 &  -  & W3 \\
22 & 20:21:10.21 & 40:30:53.78 & 0.55 & 1.64 $\pm$ 0.25 & 1.97 $\pm$ 0.26 &  -  & - \\
23 & 20:21:06.54 & 40:18:44.75 & 0.49 & 2.55 $\pm$ 0.33 & 3.23 $\pm$ 0.42 &  -  & - \\
24 & 20:21:01.74 & 40:34:42.45 & 0.38 & 5.74 $\pm$ 0.52 & 5.06 $\pm$ 0.51 &  -  & - \\
25 & 20:20:59.24 & 40:31:42.71 & 0.63 & 1.37 $\pm$ 0.26 & 1.24 $\pm$ 0.26 &  -  & - \\
26 & 20:20:57.37 & 40:28:27.62 & 0.34 & 4.49 $\pm$ 0.37 & 4.42 $\pm$ 0.40 &  -  & - \\
27 & 20:20:57.01 & 40:33:26.54 & 1.10 & 1.40 $\pm$ 0.31 & 1.11 $\pm$ 0.28 &  -  & - \\
28 & 20:20:54.62 & 40:28:21.02 & 1.10 & 0.71 $\pm$ 0.19 & 1.12 $\pm$ 0.26 &  -  & - \\
29 & 20:20:53.15 & 40:30:24.83 & 0.51 & 2.02 $\pm$ 0.31 & 2.23 $\pm$ 0.35 &  -  & - \\
30 & 20:20:52.33 & 40:24:29.23 & 0.29 & 3.04 $\pm$ 0.35 & 3.54 $\pm$ 0.49 &  -  & - \\
31 & 20:20:52.09 & 40:28:26.67 & 0.46 & 4.53 $\pm$ 0.42 & 3.61 $\pm$ 0.40 &  -  & - \\
32 & 20:20:51.11 & 40:30:31.22 & 0.84 & 1.30 $\pm$ 0.27 & 1.32 $\pm$ 0.31 &  -  & - \\
33 & 20:20:40.65 & 40:27:03.54 & 0.40 & 5.89 $\pm$ 0.53 & 4.51 $\pm$ 0.49 &  -  & - \\
34 & 20:20:33.41 & 40:18:25.60 & 0.69 & 5.31 $\pm$ 0.82 & 6.74 $\pm$ 0.92 &  -  & - \\
35 & 20:22:00.96 & 40:33:41.04 & 0.89 &  -  & 1.33  0.28 &  -  & - \\
36 & 20:21:55.92 & 40:38:31.92 & 0.85 &  -  & 0.84  0.30 &  -  & - \\
37 & 20:21:51.36 & 40:32:30.48 & 1.31 &  -  & 1.48  0.25 &  -  & - \\
38 & 20:21:37.44 & 40:36:40.68 & 2.40 &  -  & 1.10  0.31 &  -  & - \\
39 & 20:21:25.44 & 40:36:14.40 & 1.53 &  -  & 1.33  0.30 &  -  & - \\
40 & 20:21:25.20 & 40:38:15.00 & 0.78 &  -  & 1.37  0.30 &  -  & - \\
41 & 20:21:20.16 & 40:34:55.92 & 1.28 &  -  & 0.68  0.20 &  -  & - \\
42 & 20:21:07.20 & 40:38:53.52 & 0.74 &  -  & 1.29  0.35 &  -  & - \\
\hline
\hline
\multicolumn{8}{l}{\footnotesize (a) c.f Fig.~\ref{xmm_color}.}\\
\multicolumn{8}{l}{\footnotesize (b) $1\sigma$ positional uncertainty.}\\
\multicolumn{8}{l}{\footnotesize (c) W$x$ indicates the source detected independently by Weisskopf et al. (2011) with 
$x$ corresponds to the label in their Fig.~1.}
\end{longtable}
\end{landscape}

\appendix
\section{Analysis of the X-ray flash-like event XMM~J202154.7+402855} 
\begin{figure}[t]
\centerline{\psfig{figure=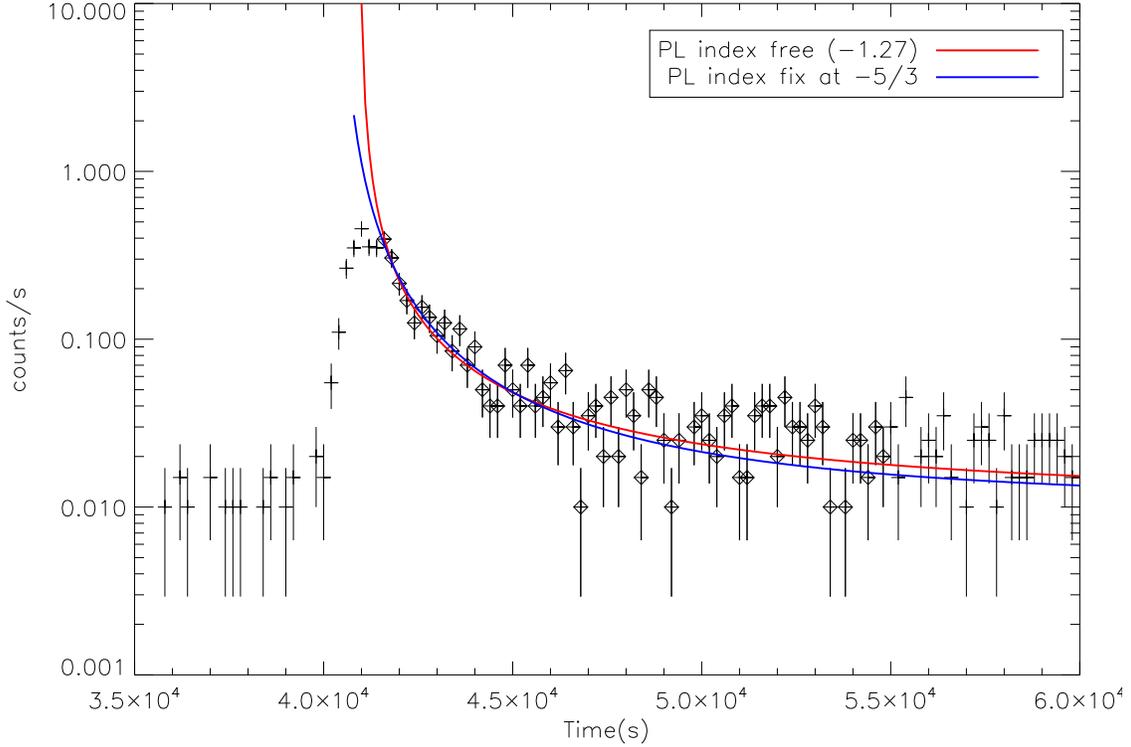, width=15cm,clip=}}
\caption[Light curve of X-ray flash]{A flash light curve of XMM J202154.7+402855 obtained from MOS 1/2. The epoch zero of the merged light curve corresponds to MJD 56028.31 related to the selected GTI of {\it XMM} archive investigated in 2012. The light curve of XMM J202154.7+402855 was binned with 200 s labelled by cross signs. The highest peak reaches to $\sim 0.4$~cts/s, and all the error bars of data points indicate the 1$\sigma$ uncertainty. Data points labelled by diamonds demonstrate the cooling tail of the burst, and the red and blue lines are the best fits to a power-law model with a free index and a fixed index at -5/3, respectively. All the obtained parameters are presented in Table~\ref{Tail_fit}.}
\label{flash_LC}
\end{figure}

In this \emph{XMM-Newton} observation, XMM~J202154.7+402855 is detected by MOS1/2 
(i.e. source 8 in Fig.~\ref{xmm_color} and Tab.~1). In examining its light curve, we found that this source is significantly 
variable and resembles a flash-like event (see Figure~\ref{flash_LC}). A rapid rise of intensity of XMM~J202154.7+402855 
occured at $\sim40$~ks 
after the start of the investigation at $\sim$ MJD 56028.31. Its count rate increased by a factor of $\sim40$ above the 
quiescent level with a timescale of $\sim1$~ks and became the brightest one among all the point sources detected in this observation. 
After reaching the peak, its count rate returned to the quiescent level in about an hour. In view of its interesting temporal 
behaviour, we carefully examined timing and spectral properties of this newly detected flash-like event. 

In order to probe the spectral behavior of XMM~J202154.7+402855, we divided its spectrum into two components: 
the quiescent spectrum (events in 0--35000~s \& 80000--110000~s) and the X-ray flash spectrum (events in 40000--70000~s)
\footnote{The time intervals here indicates the time after the start of the investigation}. 
For the quiescent spectrum, we have examined it with various single component model: 
power-law, blackbody and comptonization of soft photons in a hot plasma (Titarchuk 1994). None of this single component model 
can provide an acceptable description of the data. We proceeded to fit the quiescent spectrum with composite models. 
We found that BB+PL can fit the data reasonably well ($\chi^2$=16.3  with 12 d.o.f.), which yields a column density of 
$N_{H}=6.6^{+1.0}_{-1.1}\times10^{21}$ cm$^{-2}$, a photon index of $\Gamma=1.8^{+1.3}_{-1.4}$, a PL normalization at 1 keV of 
$5.3^{+1.4}_{-1.5}\times10^{-6}$ photons~keV$^{-1}$~cm$^{-2}$~s$^{-1}$, a blackbody temperature of $kT=97.2^{+4.3}_{-5.5}$~eV and 
an emitting radius of $R=7.8^{+2.4}_{-2.9}d_{2}$~km. 
The unabsorbed flux of XMM~J202154.7+402855 in the quiescent state is $\sim9\times10^{-13}$~erg~cm$^{-2}$~s$^{-1}$ in 0.5-10 keV.

For investigating the X-ray flash spectrum, we further divided it into three stages: a rising stage 
(events in 40000--41400~s); a rapid declining stage (events in 41400-45000~s) and a slow declining stage (events in 45000-70000~s).
For accounting the quiescent contribution in all these segments, we included the BB and PL components in the spectral fits with 
the parameters fixed at the best-fit values of the quiescent spectrum. In view of the narrow time-windows in dividing these stage, 
their photon statistics are lower than the quiescent spectrum. 
In order to minimize the number of free parameters, we also fixed the column absorption
at $N_{\rm{H}}=6.6\times 10^{21}$ $\hbox{cm}^{-2}$ as inferred from the quiescent spectrum. On top of the quiescent level, 
we have added an extra component for modeling the contribution from the flash-like event. We found that 
an additional comptonized blackbody model (Nishimura et al. 1986; XSPEC model: COMPBB) is capable to yield reasonable spectral 
fits for all three stages. With the electron temperature of the plasma fixed at 50~keV, the best-fit temperature of the comptonized blackbody 
for the rising stage, the rapid declining stage and the slow declining stage are found to be $86.9^{+21.0}_{-24.1}$~eV, 
$103.5^{+13.4}_{-12.8}$~eV and $95.9^{+15.3}_{-14.1}$~eV respectively. The sum of the unabsorbed fluxes in all three stages is 
$\sim4\times10^{-11}$~erg~cm$^{-2}$~s$^{-1}$ in 0.5-10~keV. 

Since both burst duration and spectral properties of XMM~J202154.7+402855 are inconsistent with a Type-I X-ray burst, we proceeded to 
consider other possible emission scenarios. 
We have also explored its temporal behaviour by fitting a power-law model to the fading tail of the event.
The results are summarized in Table~\ref{Tail_fit}. With all the parameters to be free, the light curve fitting yielded 
a power-law index of $-1.27\pm0.16$ and the best-fit function is shown as the red curve in Fig.~\ref{flash_LC}. 
Within its $3\sigma$ uncertainty, this value is consistent with that of a tidal disruption event (TDE) which has a time dependence 
of $t^{-5/3}$ (Rees 1988; Phinney 1989; Lodato 2012). With the power-law index fixed at -5/3, the best-fit curve (i.e. blue curve in 
Fig.~\ref{flash_LC}) also results in a comparable goodness-of-fit. 

Considering the possibility of a TDE, we further attempted to 
search for the quasi-periodic oscillation (QPO) from the data which has been detected from Swift~J1644+57 (Reis et al. 2012). 
A $\sim200$~s QPO has been detected from Swift~J1644+57 which is interpreted as the Keplerian frequency of the innermost stable 
circular orbit of a supermassive black hole (Reis et al. 2012). Motivated by this discovery, we searched for the periodic signal 
from XMM~J202154.7+402855 in a range from 0--400~s centered at 200 s with a resolution of 0.1 s by $\chi^{2}$~test. 
The highest peak obtained from the periodogram is at 115.1 s with $\chi^2_{15}$ less than 3.
Even we only considered those events obtained between the main outburst of 40000--55000 s shown in Fig.~\ref{flash_LC}, 
the similar result of the $\chi^2_{15}=3.2$ was detected at a trial period of 115 s.
We also considered to detect the periodic signal using the Lomb-Scargle method (Scargle 1982) on light curves rebinned 
with 20 s and 40 s. With this method, we concluded that there is no periodicity can be detected with a power more significant than 
90\% confidence level. Our analysis indicates that there is no stable periodic signal can be detected from the current observation.

Since QPO might appear intermittently or varies with time, these make the aforementioned periodicity search for the whole light curve 
difficult. In view of this, we have also searched the possible periodic signal by computing the dynamic power spectrum (Clarkson et al. 2003a,b). 
We adopted a window size of 1000~s which is approximately the duration of the flash-like event. 
In order to depress the effect of the trend, we used the empirical mode decompositoin (Huang et al. 1998) to filter the trend and only the 
de-trended light curve was examined by the dynamical power spectrum.
However, there is no significant signal of QPO can be detected from the dynamic Lomb-Scargle periodogram.
Together with the non-detection of any periodic signal, the fact that the burst duration of XMM~J202154.7+402855 
is far shorter than a typical TDE which lasts for a timescale of months (e.g. Reis et al. 2012) does not favor this scenario. 

Another possible source nature of XMM~J202154.7+402855 is a flaring early-type star. This requires a search for the optical counterpart for
constraining its properties. Utilizing the USNO-B1.0 catalog (Monet et al. 2003), we have identified a bright source, 
USNO-B1.0~1304-0388936, locates at $\sim3"$ away from the nominal X-ray position of XMM~J202154.7+402855 with magnitudes of
$B=14.48$, $R=13.65$ and $I=12.85$. Assuming it is the optical counterpart of XMM~J202154.7+402855, its X-ray-to-optical flux ratio 
is $f_{x}/f_{opt}\sim10^{-3}$ during quiescence which is quite typical for a field star (Maccacaro et al. 1988). To investigate if 
the positional offset between the X-ray source and the optical counterpart is a result of proper motion, 
we have also checked the UCAC3 catalog (Zacharias et al. 2010). The source 
3UC~261-199420 in UCAC3 has its nominal position differed from USNO-B1.0~1304-0388936 only by $\sim0.5"$ and its proper motion  
is very small: $\mu_{\rm RA}=1.9$~mas/yr and $\mu_{\rm Dec}=-0.9$~mas/yr. Calculating the angular shift from the central epoch for 
the position given by UCAC3 to the epoch of our \emph{XMM-Newton} observation, we found that it only moves by $\sim31$~mas.
Hence, the offset between 
XMM~J202154.7+402855 and the optical source cannot be reconciled by the proper motion. 
Although the positional offset is comparable to the absolute point accuracy of \emph{XMM-Newton}
\footnote{see http://xmm.esac.esa.int/external/xmm\_user\_support/documentation/uhb\_2.1/node108.html}, 
further investigation is required for securing the association between XMM~J202154.7+402855 and the optical source. 

\begin{table}
\caption{Best-fit parameters with both free/fixed power-law indices in $y=a_1+a_2 \times(t-a_3)^{a_4}$ to the tail of X-ray flash}\label{Tail_fit}
\begin{tabular}{lccccc}
\hline \hline Parameters & $a_1$ (cts/s) & $a_2$ & $a_3$ (s) & $a_4$ & $\chi^2_{\nu}$ $(d.o.f)$
\\
\hline\hline
$a_4$ free & 0.01 & 1511 $\pm$ 2074 & 40950 $\pm$ 215 & -1.27 $\pm$ 0.16 & 1.099(62)
\\ $a_4$ fixed & 0.01 & 49112 $\pm$ 5128 & 40388 $\pm$ 121 & -5/3 & 1.117(63)
\\
\hline\hline
\end{tabular}\end{table}

\end{document}